\documentclass[reprint, superscriptaddress, nofootinbib, altaffilsymbol, aps, prc, floatfix]{revtex4-1}
\usepackage[colorlinks=true,pdfstartview=FitV,linkcolor=blue,citecolor=blue,urlcolor=blue,bookmarks=false]{hyperref}
\usepackage[usenames,dvipsnames]{color}
\usepackage[sc]{mathpazo}
\usepackage[below]{placeins}
\usepackage{afterpage}
\usepackage[separate-uncertainty,retain-explicit-plus, per-mode = symbol, detect-all=true, range-units=brackets, tight-spacing=true]{siunitx}
\usepackage{multirow}
\usepackage{rotating}
\usepackage[displaymath, mathlines]{lineno}
\usepackage{graphicx}
\usepackage{paralist}
\usepackage{wrapfig}
\usepackage{appendix}
\usepackage{vruler}
\usepackage{etoolbox}
\usepackage{tikz}
\usepackage{listings}
\usepackage{amsmath}
\usepackage{relsize}
\usepackage{etoolbox}
\usepackage{makecell}
\usepackage{gensymb}

\modulolinenumbers[5]
\newcommand{\trit}{$^3$H}

\newcommand{\sod}{$^{22}$Na}

\newcommand{\isotope}[2]{$^{#2}${#1}}

\newcommand{\cosine}{\mbox{COSINE-100}}

\newcommand{\note}[1]{{\color{purple}#1}}

\DeclareSIUnit\kgsi{kg_\text{Si}}
\DeclareSIUnit\bqkg{Bq/\kgsi}
\DeclareSIUnit\atom{atom}
\DeclareSIUnit\atoms{atoms}
\DeclareSIUnit\atomstrit{atoms (^{3}H)}
\DeclareSIUnit\atomsber{atoms (^{7}Be)}
\DeclareSIUnit\atomssod{atoms (^{22}Na)}
\DeclareSIUnit\neutrons{neutrons}
\DeclareSIUnit\muons{\micro^-}
\DeclareSIUnit\gpg{g/g}
\DeclareSIUnit\c{$c$}
\DeclareSIUnit\second{s}
\DeclareSIUnit\day{day}
\DeclareSIUnit\dayshort{d}
\DeclareSIUnit\week{w}
\DeclareSIUnit\year{yr}
\DeclareSIUnit\standard{std}
\DeclareSIUnit\str{sr}
\DeclareSIUnit\eqquanta{eq.q.}
\DeclareSIUnit\decays{decays}
\DeclareSIUnit\inch{inch}
\newif\ifcolorfigs
\colorfigstrue        

\usepackage{subfigure}
\begin{document}
\title{Cosmogenic activation of sodium iodide}


\newcommand{\pnnl}{Pacific Northwest National Laboratory, Richland, WA 99352, USA}
\newcommand{\lanl}{Los Alamos National Laboratory, Los Alamos, New Mexico 87545, USA}
\newcommand{\anu}{Department of Nuclear Physics and Accelerator Applications, Research School of Physics, The Australian National University, Canberra, ACT 2601, Australia}
\newcommand{\ua}{Department of Physics and Astronomy, University of Alabama, Tuscaloosa, Alabama 35487, USA}
\newcommand{\yale}{Department of Physics and Wright Laboratory, Yale University, New Haven, Connecticut 06520, USA}
\newcommand{\ucsd}{Department of Physics, University of California San Diego, La Jolla, California 92093, USA}
\newcommand{\CDMPP}{ARC Centre of Excellence for Dark Matter Particle Physics, Australia}
\author{R.~Saldanha}\email[Corresponding author: ]{richard.saldanha@pnnl.gov}\affiliation{\pnnl}
\author{W.G.~Thompson}\affiliation{\yale}
\author{Y.Y.~Zhong}\affiliation{\anu}\affiliation{\CDMPP}
\author{L.J.~Bignell}\affiliation{\anu}\affiliation{\CDMPP}
\author{R.H.M.~Tsang}\affiliation{\ua}
\author{S.J.~Hollick}\affiliation{\yale}
\author{S.R.~Elliott}\affiliation{\lanl}
\author{G.J.~Lane}\affiliation{\anu}\affiliation{\CDMPP}
\author{R.H.~Maruyama}\affiliation{\yale}
\author{L.~Yang}\affiliation{\ucsd}


\begin{abstract}
The production of radioactive isotopes by interactions of cosmic-ray particles with sodium iodide (NaI) crystals can produce radioactive backgrounds in detectors used to search for rare events. Through controlled irradiation of NaI crystals with a neutron beam that matches the cosmic-ray neutron spectrum, followed by direct counting and fitting the resulting spectrum across a broad range of energies, we determined the integrated production rate of several long-lived radioisotopes. The measurements were then extrapolated to determine the sea-level cosmogenic neutron activation rate, including the first experimental determination of the tritium production rate: \mbox{$(80 \pm 21)$ atoms/kg/day}. These results will help constrain background estimates and determine the maximum time that NaI-based detectors can remain unshielded above ground before cosmogenic backgrounds impact the sensitivity of next-generation experiments.
\end{abstract}


\maketitle

\section{Introduction}
\label{sec:intro}

Thallium-doped sodium iodide (NaI(Tl)\footnote{hereafter referred to simply as NaI}) detectors have been at the forefront of nuclear and particle physics research for nearly 75 years \cite{knoll}. Because of their relatively high light yield, general ease of use, and availability at a relatively low cost, NaI detectors still enjoy widespread use, despite the development of newer radiation detection technologies. One prominent application of NaI detectors is in the field of dark matter direct detection.

Of particular note, the DAMA/LIBRA dark matter experiment \cite{DAMA_2018} comprises a 250 kg array of ultra-low-background NaI detectors. For nearly two decades, the DAMA collaboration has claimed a detection of dark matter in the form of an annual modulation of the event rate in their NaI crystals \cite{DAMA_2018}. This observed modulation has a period and phase that are consistent with a dark matter-induced modulation signal as predicted by the standard halo model \cite{Drukier_1986}. In addition, DAMA's observed modulation signal has persisted for over 20 years and is now observed at a significance of 13.7$\sigma$ compared with the no-modulation hypothesis \cite{DAMA_2021}.

Despite the high confidence at which DAMA observes this modulation, non-NaI-based direct detection experiments have obtained null results in their searches for dark matter, casting doubt on DAMA's claim of dark matter discovery. This tension in the field has given rise to several NaI-based direct detection experiments that aim to perform a model-independent test of DAMA's claim of dark matter discovery by using the same target material. Three of these experiments, \mbox{DM-Ice} \cite{dmice_prd}, \cosine~\cite{cosine_3yr,thesis_thompson}, and ANAIS-112 \cite{anais3yrmodulation}, are currently running and have performed initial tests of the DAMA claim, while a fourth, SABRE \cite{Antonello_2019, SABRE_2022} is currently under construction.

A significant challenge for these NaI-based experiments is the design and manufacture of NaI detectors with background activity levels comparable to that of DAMA/LIBRA, which utilizes some of the lowest-background NaI crystals ever produced. This has led to significant research and development efforts by both the COSINE and SABRE collaborations to develop ultra-low background detectors \cite{cosine_det,Antonello_2021}. Much of this R\&D work has focused on reducing contamination of radioactive impurities, primarily \isotope{Pb}{210} and \isotope{K}{40}, that are introduced into the crystal during the growth and encapsulation stages.

Given the recent successes of these background reduction efforts, it is likely that in next-generation NaI dark matter experiments cosmogenic radioisotopes will form the primary background component below 10 keV, the region of interest for dark matter searches. Tritium, a pure $\beta$-emitter with an end point at 18.6 keV, is of particular interest because of its apparent high cosmogenic activity in the target NaI crystals, as seen in ANAIS-112 \cite{anais2018activationrates} and COSINE-100 \cite{COSINE100activationrates}. Additionally, \isotope{Na}{22}, with a K$_\alpha$ x-ray line at 0.85 keV, is expected to be a significant background source, as future NaI experiments aim to push to energy thresholds below $\sim$\SI{1}{\keV}.

In order to meet the ultra-low background goals of these future experiments, it will be necessary to limit the amount of time that individual NaI detectors remain unshielded from cosmic rays, particularly neutrons, which can produce radioactive isotopes within the crystal. The level of cosmogenic activation of a particular isotope is effectively determined by the above-ground exposure time, the cosmic ray flux, and the production cross section of the isotope. For many isotopes of interest there are few direct measurements of the production cross sections. While several estimates of the cross sections based on semi-empirical calculations and nuclear models exist, these estimates can vary significantly, as seen in Fig.~\ref{fig:cs_trit}, leading to a large uncertainty in the acceptable above-ground residency time. This has led sodium iodide-based dark matter experiments, including ANAIS~\cite{Amare:2014bea,anais2018activationrates,Amare:2018ndh}, \mbox{DM-Ice}~\cite{dmice17activationrates} and \cosine~\cite{COSINE100activationrates}, to calculate production rates for isotopes using the measured isotopic activities and the estimated above-ground exposure history of a given NaI detector.

In this paper, we present results from a dedicated measurement of cosmogenic isotope production rates that utilizes NaI crystals activated in a neutron beam with a spectrum that approximates the cosmic ray spectrum. Our approach allows a precise knowledge of the exposure history of the crystal. This mitigates the main source of systematic uncertainty present in previous studies, at the cost of a new source of systematic uncertainty associated with the difference between the beam conditions and true cosmic ray exposure. Our irradiations were performed at the Los Alamos Neutron Science Center (LANSCE) ICE-HOUSE II facility \cite{lisowski2006alamos, icehouse}. The ICE-HOUSE II neutron beam has a similar energy spectrum to that of cosmic ray neutrons, but with a flux $\sim5 \times 10^8$ times larger than the natural sea-level flux. This facility is well-suited for cosmogenic activation studies and has previously been used to measure cosmogenic activation cross-sections for argon \cite{Saldanha2019-Ar} and silicon \cite{Saldanha2020-Si} targets. The high beam flux allows for the detection of measurable amounts of cosmogenic radioisotopes in NaI detectors in an above-ground facility with an exposure of just a few hours. After irradiation at the ICE-HOUSE II facility, we measured the isotope decay rates in the activated detectors and extrapolated this measurement to determine the integrated production rate by cosmic rays of several isotopes of interest for dark matter searches, including the first measurement of the production rate of \isotope{H}{3}. This measurement will enable an accurate determination of acceptable above-ground residency times for sodium iodide detectors and help constrain the contributions of activation products to the overall background rates, helping future NaI-based dark matter experiments to meet their background goals. 
\section{Cosmogenic Radioisotopes}
\label{sec:isotopes}
\begin{table}[t]
\centering
\begin{tabular}{c c c c}
\hline
Isotope & Half-Life & Decay & Q-value \\
 & [d] & Mode & [keV]\\
\hline
\vrule width 0pt height 2.2ex
$^3$H & \num{4500 \pm 7} & $\beta$- & \num{18.591 \pm 0.003} \\
$^{22}$Na & \num{950.4 \pm 0.7} & $\beta$+ & \num{2842.2 \pm 0.2}\\
$^{109}$Cd  & \num{461.9 \pm 0.4} & EC & \num{215.1 \pm 2.0} \\
$^{109m}$Ag & \num{4.60 \pm 0.02 E-4} & IT & \num{88.034 \pm 0.001} \\
$^{113}$Sn & \num{115.09 \pm 0.03} & EC & \num{1039 \pm 2} \\
$^{113m}$In & \num{6.908 \pm 0.02 E-2} & IT & \num{391.699 \pm 0.003} \\
$^{121m}$Te & \num{164.4 \pm 0.7} & IT & \num{293.974 \pm 0.022} \\
$^{121}$Te & \num{19.17 \pm 0.04} & EC & \num{1056 \pm 26} \\
$^{123m}$Te & \num{119.2 \pm 0.1} & IT & \num{247.45 \pm 0.04} \\
$^{125m}$Te & \num{57.4 \pm 0.1} & IT & \num{144.775 \pm 0.008}\\
$^{127m}$Te & \num{105.9 \pm 0.7} & IT & \num{88.23 \pm 0.07} \\
$^{127}$Te & \num{0.390 \pm 0.003} & $\beta$- & \num{703 \pm 4} \\
$^{125}$I & \num{59.41 \pm 0.01} & EC & \num{185.77 \pm 0.06} \\
\hline
\end{tabular}
\caption{List of all radioisotopes with half-lives $>$\,30 days, as well as their progeny, which can be produced by cosmogenic interactions with NaI(Tl) and were considered in this work. For reference, we also list the half-life, primary decay mode, and Q-value. All data are taken from NNDC databases \cite{dunford1998online}.}
\label{tab:rad_isotopes}
\end{table}

Interactions of high energy cosmogenic particles with NaI crystals can produce a large number of radioisotopes, in principle any isotope lighter than the target isotopes. The production rates are largest for isotopes close to the target isotopes ($^{23}$Na and $^{127}$I) and for light isotopes such as $^3$H which can be ejected from the impacted nucleus. For dark matter experiments the most dangerous isotopes are those that are relatively long-lived and whose decay chain produces interactions in the NaI crystal that overlap in energy with the expected dark matter signal. In Table~\ref{tab:rad_isotopes} we have listed the radioisotopes with half-lives longer than a month (and any radioactive progeny) that were considered. 

At sea-level the production rate of isotopes is dominated by interactions induced by high-energy neutrons, with interactions of protons typically contributing $\lesssim 10\%$ and even smaller contributions from muons and gammas \cite{anais2018activationrates}. The neutron-induced production cross-sections for nearly all of these isotopes have not been directly measured, with the exception of $^3$H, $^{22}$Na, and $^{125}$I, which we discuss in detail below.
\subsection{$^3$H}
\begin{figure}
    \centering
    \includegraphics[width=\columnwidth]{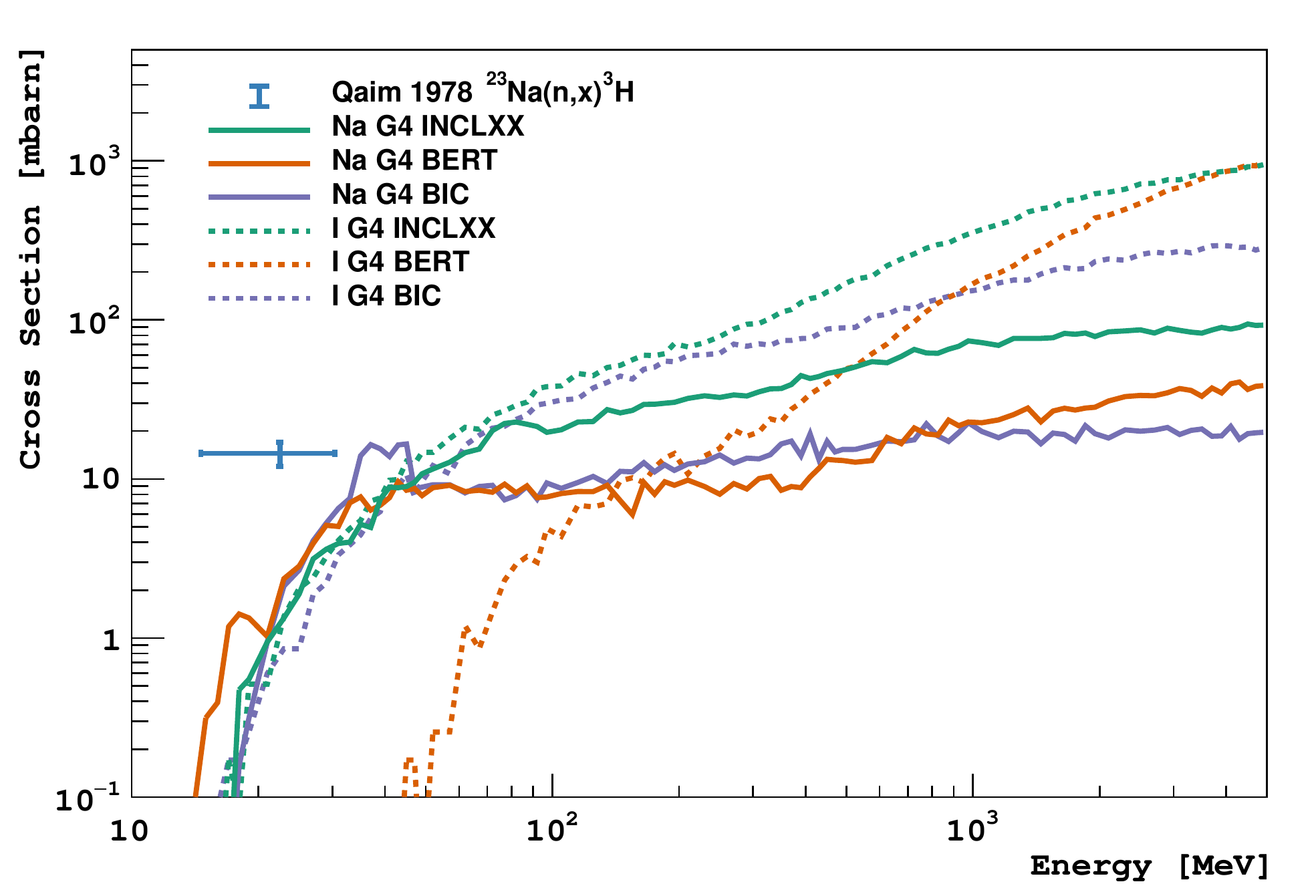}
    \caption{Experimental measurement \cite{qaim1978triton} and Geant4 model estimates (see text for details) of neutron-induced tritium production in sodium (solid curves) and iodine (dotted curves).}
    \label{fig:cs_trit}
\end{figure}
Tritium is a particularly dangerous background for dark matter searches because it decays by pure beta emission, and its low Q-value (18.6 keV) results in a large fraction of decays that produce low-energy events in the expected dark matter signal region. There is only a single measurement of the production cross-section of tritium from $^{23}$Na at relatively low energies \cite{qaim1978triton} and to the best of our knowledge there are no integrated production rate measurements from dark matter experiments. Figure~\ref{fig:cs_trit} shows the single experimental measurement and predictions for the cross-section on both $^{23}$Na and $^{127}$I based on the models built into Geant4\footnote{We used Geant4.10.3.p02 with physics lists QGSP\_INCLXX 1.0 (INCL++ v5.3) \cite{boudard2013new,mancusi2014extension}, QGSP\_BERT 4.0 \cite{bertini1963low, guthrie1968calculation, bertini1969intranuclear, bertini1971news}, and QGSP\_BIC 4.0 \cite{folger2004binary}.} \cite{allison2016recent, agostinelli2003geant4}. The Geant4 cross-sections, shown in Figure~\ref{fig:cs_trit} and subsequent figures, were extracted by targeting neutrons at various energies towards a 1 mm thick NaI target and calculating the fraction of primary neutron events in which the relevant activation products were created.

\subsection{$^{22}$Na}
\begin{figure}
    \centering
    \includegraphics[width=\columnwidth]{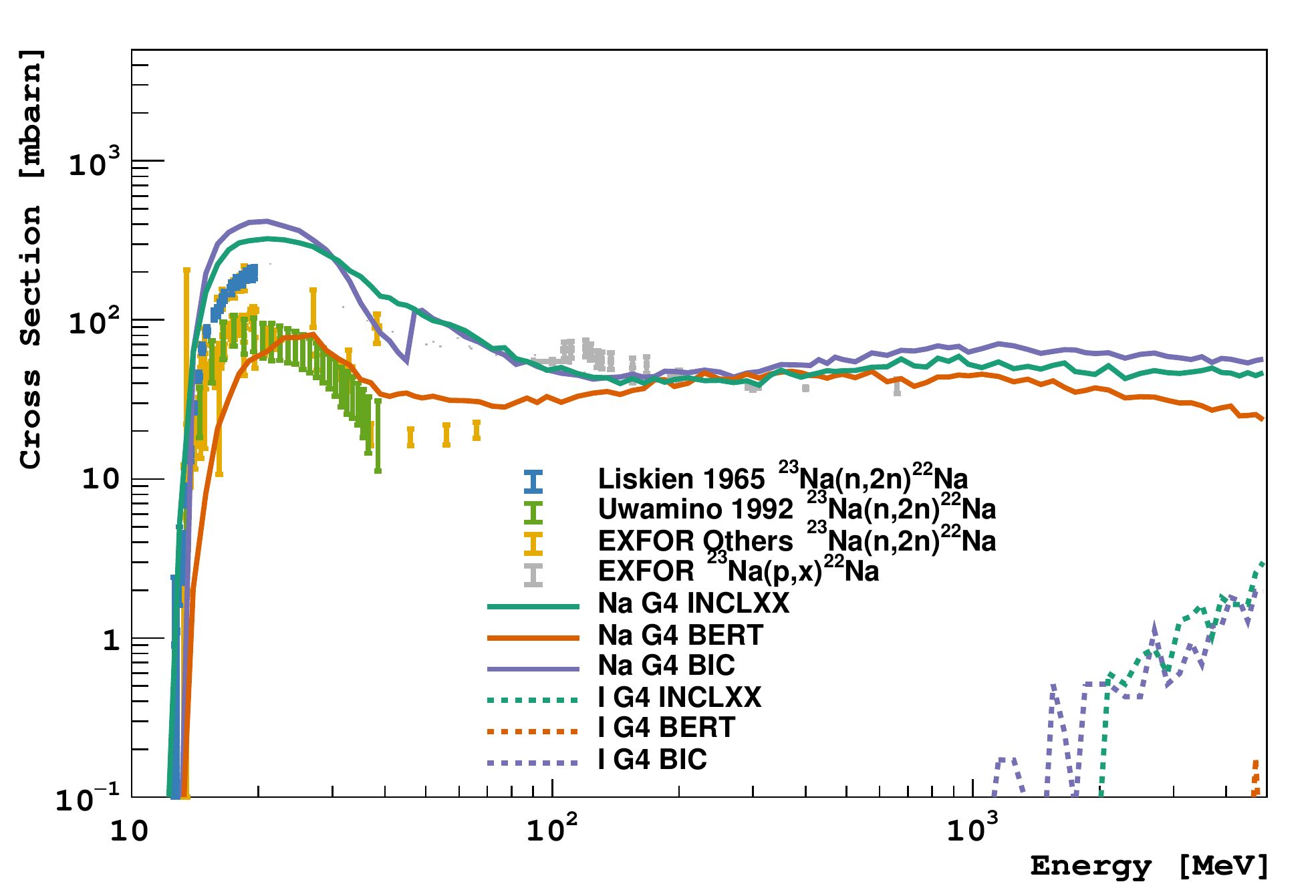}
    \caption{Experimental measurements \cite{exfor, liskien1965cross, uwamino1992measurement} and Geant4 model estimates of neutron-induced \sod~production in sodium (continuous curves) and iodine (dotted curves). Measurements of the proton-induced cross section (grey markers) \cite{exfor} are also shown for reference.}
    \label{fig:cs_na}
\end{figure}
\sod~decays primarily by positron emission (90.3\%) or electron capture (9.6\%) to the 1275 keV level of $^{22}$Ne. \sod~can be an important background as it is possible that both the \SI{1275}{\keV} $\gamma$ ray and the \SI{511}{\keV} positron-annihilation photons will escape undetected, with only the emitted positron (end-point \SI{547}{keV}) or atomic de-excitation following electron capture ($\sim$ \SI{0.85}{keV}) depositing energy in the crystal. Due to the relatively high production rate and characteristic gamma rays emitted during the decay of $^{22}$Na, there are several measurements of the neutron-induced production cross-section below 100 MeV, as shown in Figure~\ref{fig:cs_na}. It can be seen that near the expected peak in cross-section there is disagreement between the two most extensive sets of measurements from Liskien and Paulsen \cite{liskien1965cross} and Uwamino et. al. \cite{uwamino1992measurement}. For reference we also show measurements of proton-induced cross-section \cite{exfor}, which should be similar to the neutron-induced cross-section at high energies ($>$ \SI{100}{MeV}).
\subsection{$^{125}$I}
\begin{figure}
    \centering
    \includegraphics[width=\columnwidth]{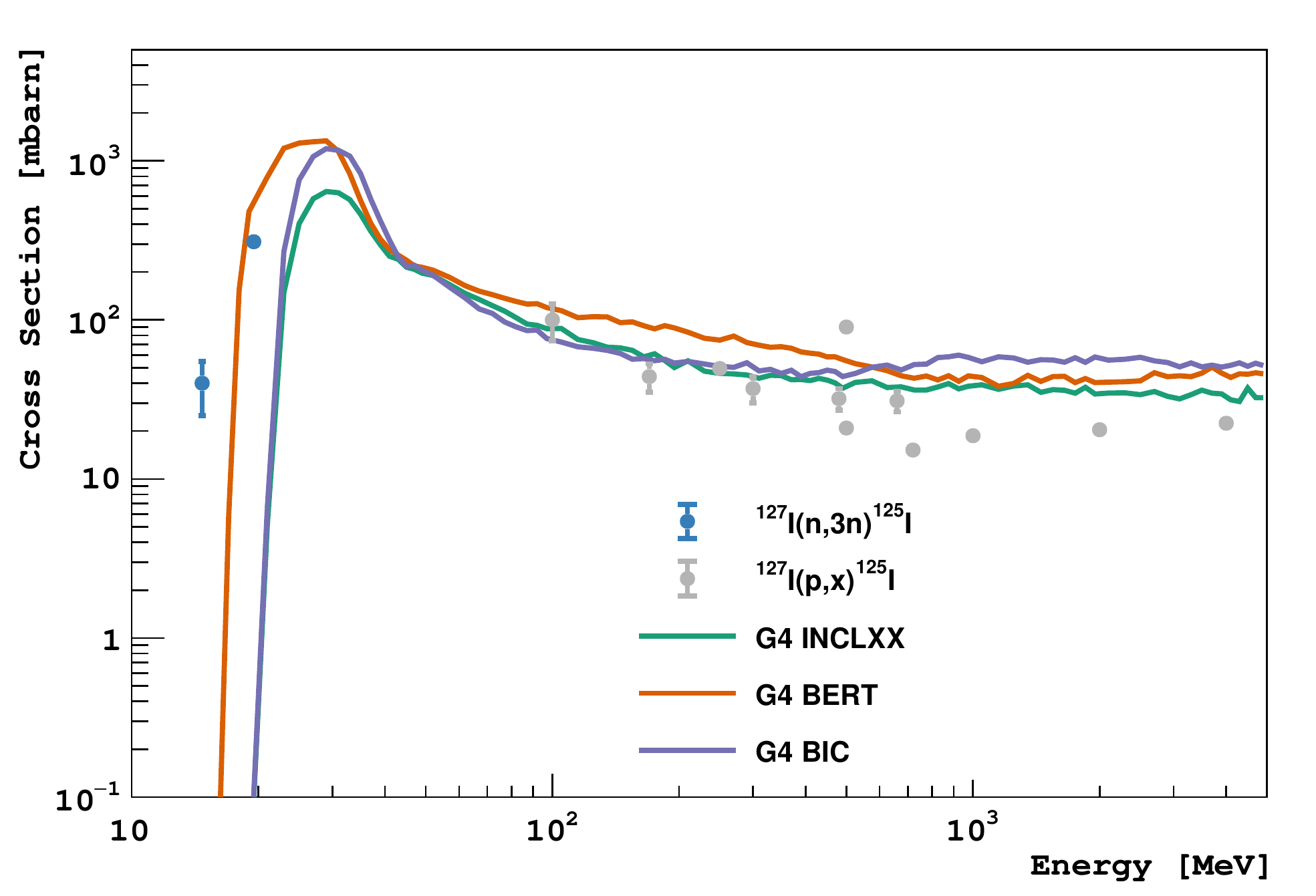}
    \caption{Experimental measurements \cite{qaim1968half, liskien1968n} and Geant4 model estimates (continuous curves) of neutron-induced $^{125}$I production in iodine. Measurements of the proton-induced cross section (grey markers) \cite{exfor} are also shown for reference.}
    \label{fig:cs_iodine}
\end{figure}
$^{125}$I decays by electron capture to the \SI{35.5}{keV} level of $^{125}$Te. $^{125}$I has a half-life of 59.4 days and so is not typically a problematic background for dark matter searches that keep the NaI crystals deep underground for much longer periods. Nevertheless its large production rate and distinctive peaks make it easy to measure and a good calibration of the neutron exposure. Figure~\ref{fig:cs_iodine} shows the existing cross-section measurements for neutrons \cite{qaim1968half, liskien1968n} and protons \cite{exfor} along with the Geant4 cross-section models. 

\section{Beam Exposure}
\label{sec:exposure}
To evaluate the production rate of cosmogenic isotopes through the interaction of high-energy neutrons, we irradiated NaI crystals at the LANSCE neutron beam facility. Following the irradiation, scintillation signals from the NaI crystals were read out to measure the beam-induced activity within the crystal. In this section we describe the details of the targets and beam exposure, while in Sec.~\ref{sec:counting} we present the measurement results.

\subsection{NaI Crystals}
\label{sec:crystals}
The NaI crystals were manufactured by Alpha Spectra Inc. \cite{alphaspectra} and were custom-designed for this measurement. The 0.2\% thallium-doped NaI crystals are cylinders with a \SI{3.000 \pm 0.020}{\inch} (\SI{7.620 \pm 0.051}{\cm}) diameter and \SI{1.000 \pm 0.031}{\inch} (\SI{2.540 \pm 0.079}{\cm}) thickness. One of the flat faces of the cylinder is optically coupled to a \SI{0.1}{\inch} (\SI{0.25}{\cm}) thick quartz window while the other faces are wrapped in a reflector and encapsulated in an aluminum body with \SI{0.020}{\inch} (\SI{0.051}{\cm}) wall thickness. A \SI{4}{\inch} OD flange surrounds the quartz window for coupling to a photosensor after the irradiation. The crystal encapsulation was designed for the neutron beam to pass through the crystal along the central axis of the cylinder and the thickness of all materials surrounding the crystal within the \SI{2}{\inch} (\SI{5.1}{\cm}) beam diameter, namely the aluminum casing and optical window, were minimized. The composition and thickness of the reflector wrap and optical coupling is proprietary, but was confidentially obtained from the company for use in the simulation of the beam exposure and the decay of radioisotopes. The crystals were specified to have better than 8.0\% energy resolution at \SI{662}{\keV}.

Radiation damage from the beam exposure can affect the scintillation light yield with a sufficiently high neutron dose. While we are not aware of any reports of radiation damage to NaI due to neutrons at energies relevant to the LANSCE beam, damage has been observed with fluences of $>10^{14}$ fast reactor neutrons \cite{Kubota1999NaIdmg}. Meanwhile, no damage was observed in NaI from 14 MeV neutrons after receiving a $4.7\times10^{11}$ neutron fluence \cite{Sudac2010NaIDmg}. Therefore, we decided to irradiate three crystals with varying fluences in order to obtain the best compromise between possible radiation damage and activation. We targeted neutron fluences of roughly $0.5\times10^{12}$, $1\times10^{12}$, and $3\times10^{12}$ neutrons for the three crystals, which was expected to give readily measurable activation while ensuring the dose fell far below that at which damage was reported. A fourth identical crystal was purchased but not activated in the LANSCE beam. It was used to measure the environmental background in the counting setup.

\subsection{LANSCE Beam}
\label{sec:lansce_beam}
The samples were irradiated at the LANSCE WNR ICE-HOUSE II facility~\cite{icehouse} on Target 4 Flight Path 30 Right (4FP30R). A broad-spectrum (0.2--800 MeV) neutron beam was produced via spallation of 800 MeV protons on a tungsten target. A \SI{2}{\inch} diameter beam collimator was used to restrict the majority of the neutrons to within the active region of the crystal. The neutron fluence was measured with $^{238}$U foils by an in-beam fission chamber~\cite{wender1993fission} placed downstream of the collimator. The beam has a pulsed time structure, which allows the incident neutron energies to be determined using the time-of-flight technique (TOF)---via a measurement between the proton beam pulse and the fission chamber signals~\cite{lisowski2006alamos,wender1993fission}.

\begin{figure}[ht]
\begin{center}
 \includegraphics[width=\columnwidth]{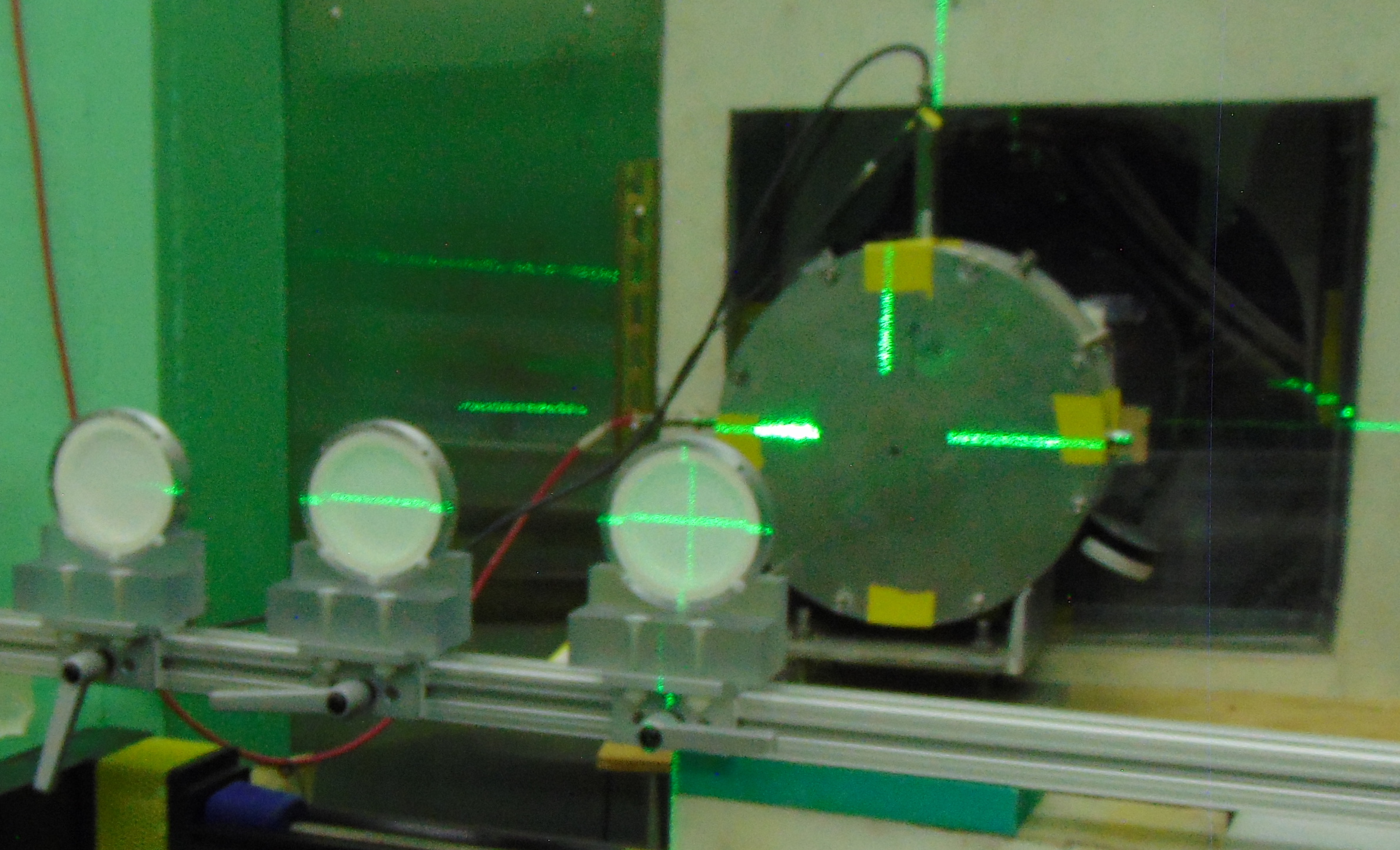}
 \end{center}
 \caption{Picture of the three NaI crystals mounted on sliding acrylic holders being aligned with the neutron beam at the LANSCE ICE-HOUSE II facility. The beam direction is angled out of the page, and passes through the cylindrical fission chamber, seen on the right, before passing through a crystal.}
 \label{fig:beamlayout}
\end{figure}

The beam exposure took place over two days between November 1$^{\mathrm{st}}$ and 3$^{\mathrm{rd}}$, 2019.  The crystals were mounted onto an acrylic holder and placed with the optical window facing away from the beam and the front face of the Al enclosure \SI{470}{\mm} from the face of the fission chamber. The center of the crystal face was aligned with the center of the beam profile using an alignment laser, as shown in Fig.~\ref{fig:beamlayout}.  Crystal A was placed in the beam line on Nov.\,1, at 19:16 local time. The crystal was periodically monitored for yellowing due to radiation damage, but no effect was visible. On Nov.\,2, at 20:17 we placed Crystal C on the beam line, replacing Crystal A. Finally, on Nov.\,3, at 04:09 we replaced Crystal C with Crystal B. The exposure was stopped at 06:30 on Nov.\,3. Following its irradiation, we measured Crystal A for 3 days using a HPGe detector starting from 20:18, Nov.\,2, to observe gamma rays emitted by the short-lived radioisotopes. The analysis of these measurements is ongoing and will be presented in a future publication. All crystals exposed to the beam were kept in storage for roughly 11 weeks to allow the radioactivity to decay down to below hazardous levels prior to shipment for the counting measurements described in Section~\ref{sec:counting}.

\subsection{Target Fluence}
\label{sec:fluence}
\begin{figure}
    \centering
    \includegraphics[width=\columnwidth]{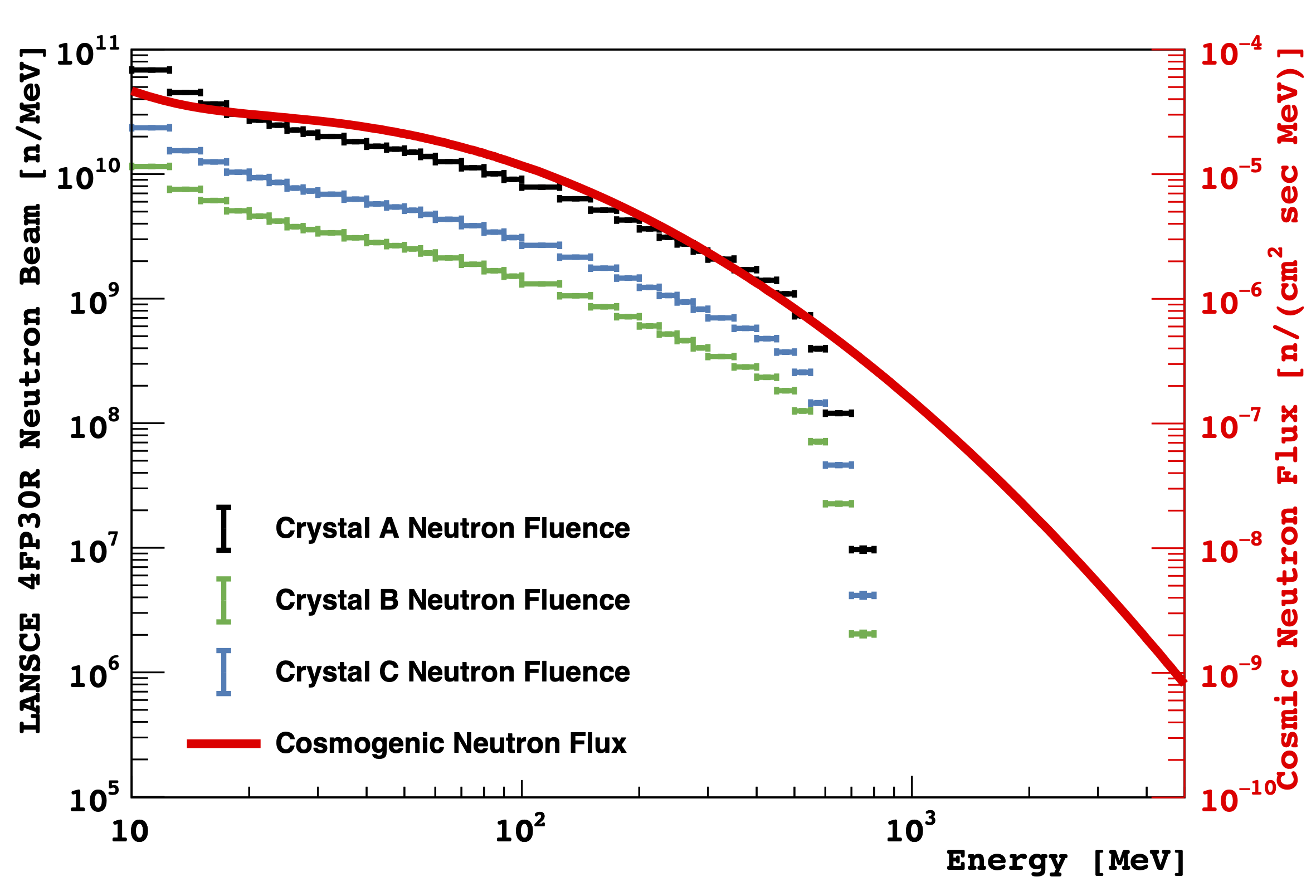}
    \caption{Comparison of the LANSCE 4FP30R/ICE II neutron beam with sea-level cosmic ray neutrons. The data points and left vertical axis show the number of neutrons measured by the fission chamber during the beam exposure for each crystal. Uncertainties shown are statistical only (see main text for discussion of systematic uncertainties). For comparison, the red continuous line and the right vertical axis show the reference cosmic-ray neutron flux at sea level for New York City during the midpoint of solar modulation \cite{gordon2004measurement}}
    \label{fig:fluence}
\end{figure}

\begin{figure}
\begin{center}
 \includegraphics[width=\columnwidth]{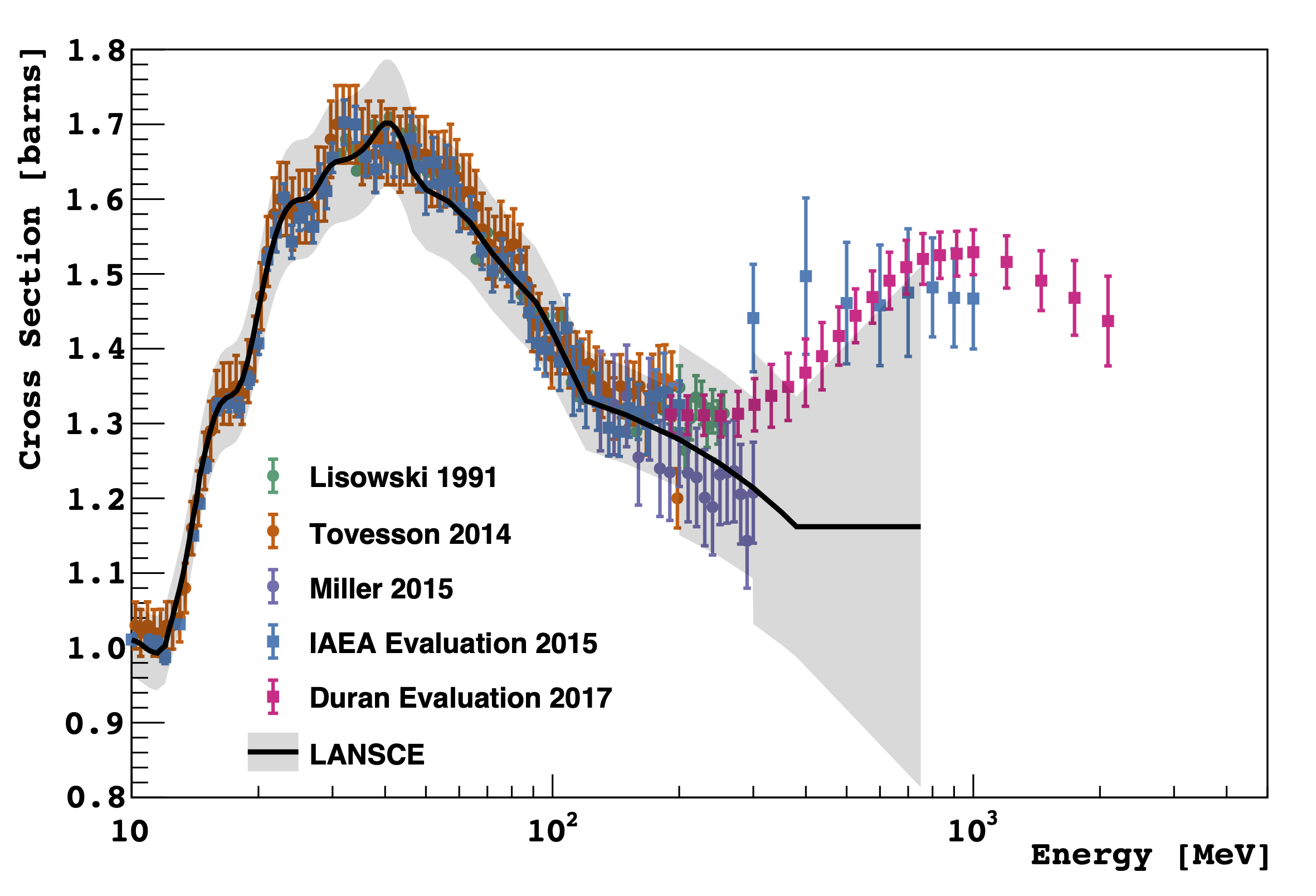}
 \end{center}
 \caption{Experimental measurements (circles) \cite{lisowski1991fission, tovesson2014fast, miller2015measurement} and evaluations (squares) \cite{carlson2009international, marcinkevicius2015209, duran2017search, carlson2018evaluation} of the $^{238}$U(n, f) cross section. The cross section assumed by the LANSCE facility to convert the fission chamber counts to a total neutron fluence is shown by the black line, with the shaded grey band indicating the assumed uncertainty (see Section~\ref{sec:fluence} for details).}
 \label{fig:fission_cs}
\end{figure}

The fluence of neutrons during each NaI crystal exposure, as measured by the fission chamber, is shown in Figure~\ref{fig:fluence}, with total fluence of \num{2.89 \pm 0.23 E12} neutrons $>$ \SI{10}{\MeV} during the Crystal A beam exposure. The uncertainty in fluence is dominated by the systematic uncertainty in the $^{238}$U(n, f) cross section used to monitor the fluence, shown in Figure~\ref{fig:fission_cs}. Below 200 MeV the assumed LANSCE cross section and various other experimental measurements and evaluations \cite{lisowski1991fission, carlson2009international, tovesson2014fast, marcinkevicius2015209} agree to better than 5\%. Between 200 and 300 MeV there are only two measurements of the cross section \cite{lisowski1991fission, miller2015measurement} which differ by 5--10\%. Above \SI{300}{\MeV} there are no experimental measurements. The cross section used by the LANSCE facility assumes a constant cross section above \SI{380}{\MeV} at roughly the same value as that measured at \SI{300}{\MeV} \cite{miller2015measurement}. This is in tension with evaluations based on extrapolations from the $^{238}$U(p, f) cross section that recommend an increasing cross section to a constant value of roughly \SI{1.5}{\barn} at 1 GeV \cite{duran2017search,carlson2018evaluation}. We have used the LANSCE cross section and assumed a 5\% systematic uncertainty below \SI{200}{\MeV}, a 10\% uncertainty between 200 and \SI{300}{\MeV}, a 15\% uncertainty between 300 and \SI{400}{\MeV}, and a constant 30\% uncertainty between 300 and \SI{750}{\MeV}. Statistical uncertainties ($\sim$ 0.3\% at the lowest energies and 3\% at the highest energies) and the uncertainty in the neutron energy spectrum due to the timing uncertainty in the TOF measurement (FWHM $\sim$ \SI{1.2}{\nano\second}) are included but are sub-dominant for this measurement.

\begin{figure}
    \centering
    \includegraphics[width=0.3\textwidth]{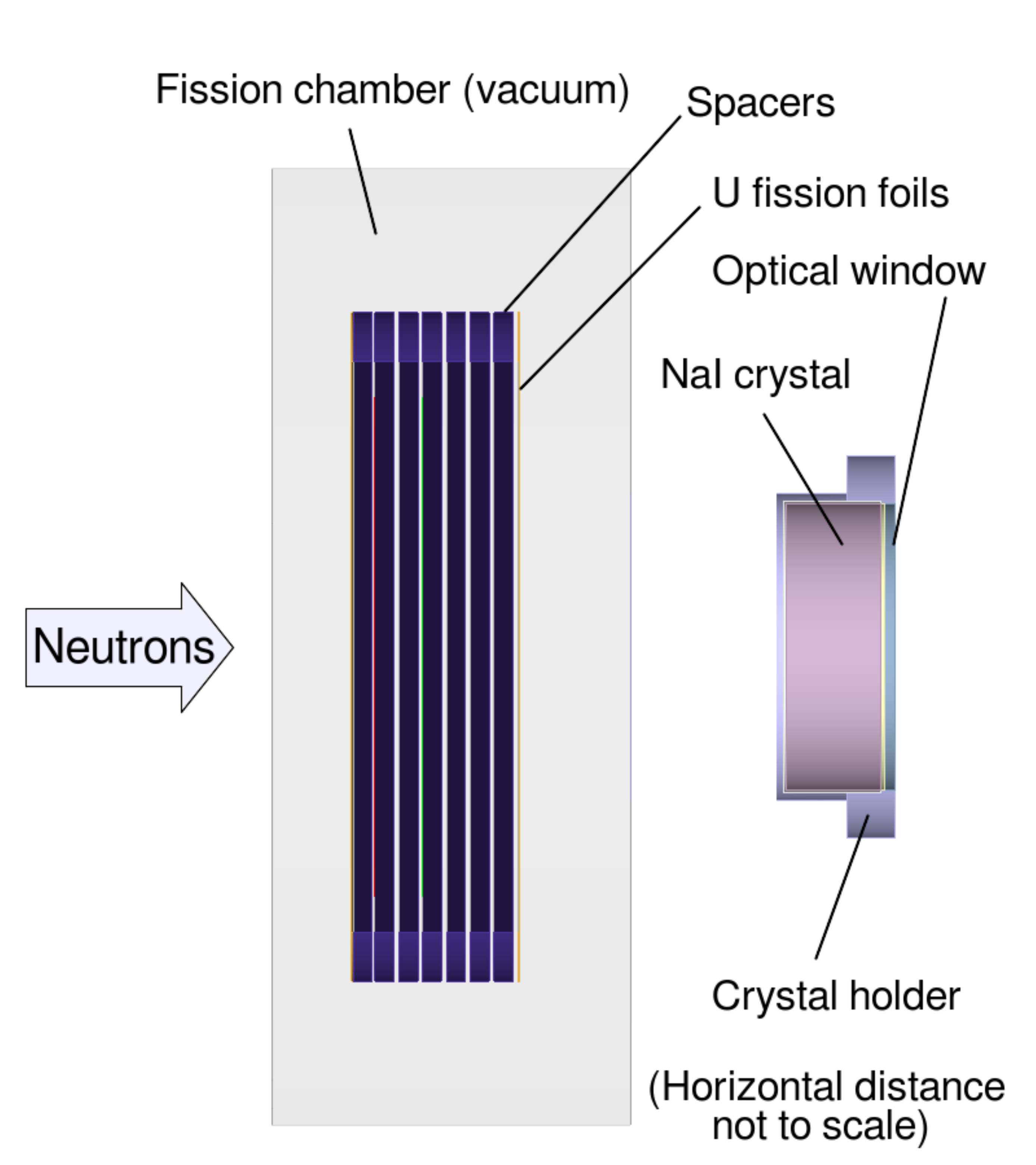}
    \caption{A Geant4 rendering showing the NaI crystal and the fission chamber on the LANSCE beam. The horizontal distance between the crystal and the fission chamber is not shown to scale due to space limitations.}
    \label{fig:g4rendering}
\end{figure}

While the nominal beam diameter was set by the \SI{2}{inch} collimator, the cross-sectional beam profile has significant tails at larger radii. At the fission chamber approximately 13\% of neutrons fall outside a \SI{3}{inch} diameter, as calculated with the beam profile provided by LANSCE. Additionally the beam is slightly diverging, with an estimated cone opening angle of 0.233\degree. A Geant4 \cite{agostinelli2003geant4,allison2016recent} simulation that included the measured beam profile and beam divergence, the measured neutron spectrum, and the full geometry (location and materials of the targets, mounting apparatus, and fission chamber~\cite{wender1993fission}), was used to calculate the neutron fluence through the NaI crystals, as illustrated in Figure \ref{fig:g4rendering}. 

Neutrons were generated uniformly in a 16 $\times$ 16 cm$^2$ square behind the fission chamber. This is sufficiently large to cover the \SI{4}{\inch} OD NaI crystal flange, the beam target component with the largest diameter. Every neutron was assigned a weight which is proportional to the intensity of the beam at the simulated neutron location, as obtained from the two-dimensional beam profile supplied by LANSCE. This allowed reuse of the same simulation results for different beam profiles and alignment offsets. A total of \num{2.4E8} neutrons above 10 MeV were simulated for each physics list such that the statistical uncertainties in the simulation are subdominant to the total neutron fluence uncertainty. To assess the uncertainty in the fluence due to misalignment of the beam with the crystal center, simulations were run with the beam shifted by the width of the alignment laser, leading to a systematic uncertainty of 0.2\%. Including all geometrical effects and systematic uncertainties, a total of \num{2.51 \pm 0.20 E12} neutrons above \SI{10}{\MeV} passed through Crystal A during the beam exposure.

The same Geant4 simulation was also used to record all radioisotopes produced in the NaI crystal and crystal holder, as predicted by the different Geant4 physics lists. The use of a full particle tracking software allowed us to include effects such as neutron attenuation and production from secondary particles in the relatively thick NaI target. The list of activation products was used to inform the species included in the decay simulations and spectral fit (described in Section~\ref{sec:analysis}), while the specific predicted activities from each of the physics lists were used to extrapolate from the beam measurement to the cosmogenic production rate (described in Section~\ref{sec:production_rates}).

\section{Counting}
\label{sec:counting}

Following the irradiation and cooling periods at LANSCE, we shipped the activated detectors offsite for measurement. Measurements of Crystal~A and Crystal~B were performed at Wright Laboratory, Yale University. Though we intended to measure Crystal~C at the Australian National University, it was lost in transit by the courier company, precluding its inclusion in this analysis. In order to quantify the activities of the neutron-activated isotopes within a detector, we measured the detector's energy spectrum through observation of the scintillation light generated by the NaI crystal. Hence, the NaI crystal functioned as both the source and detector in the measurement. As we did not observe any evidence of beam-induced radiation damage, we exclusively report measurements of Crystal~A due to its longer irradiation time and hence larger activation signal relative to environmental backgrounds. 

\subsubsection{Experimental Setup}
\begin{figure}
\begin{center}
 \includegraphics[width=\columnwidth]{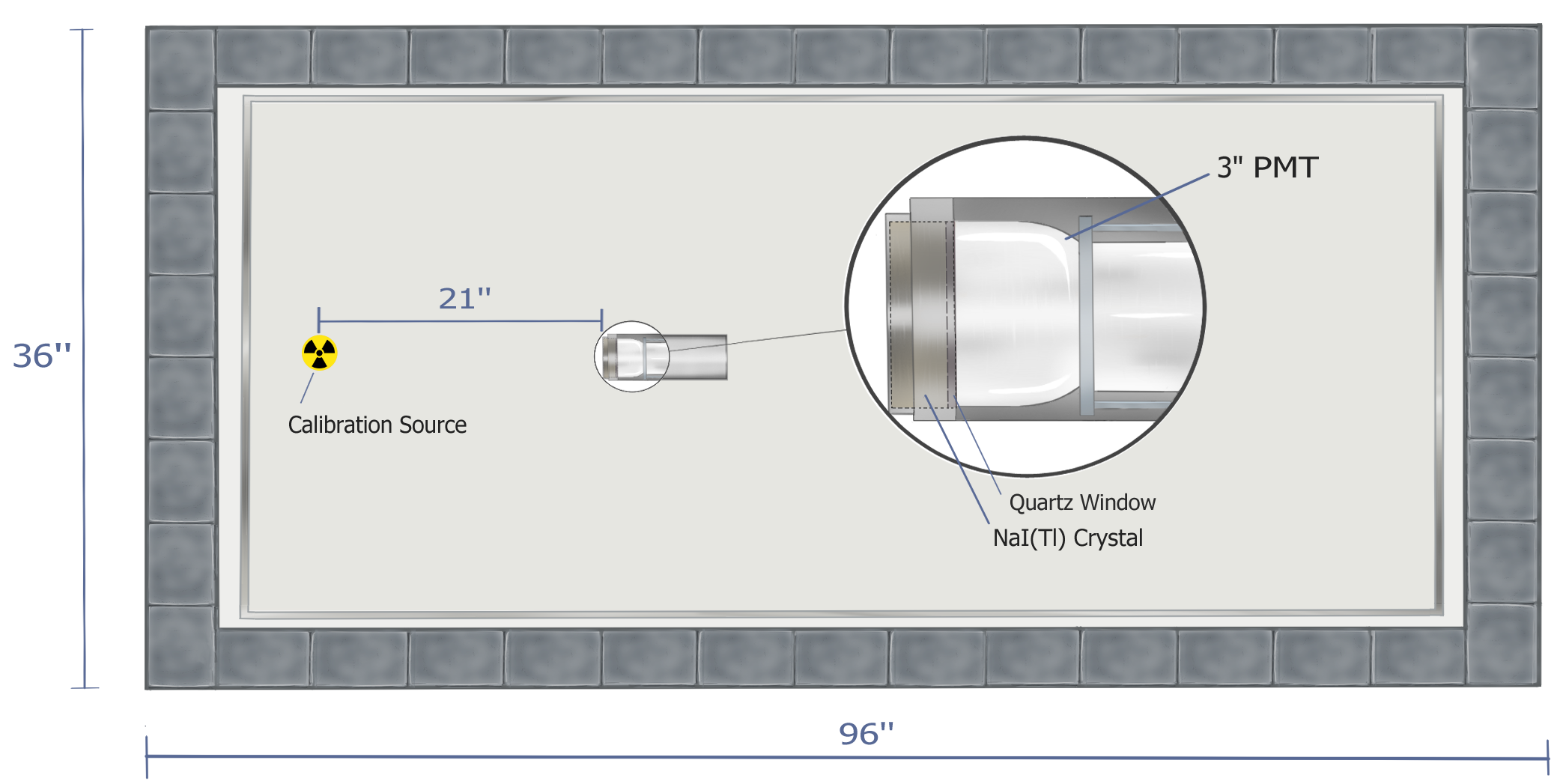}
 \end{center}
 \caption{Diagram of the experimental setup at Wright Laboratory. The irradiated NaI crystal's fused silica window was optically coupled to a PMT, which was contained in a light-tight aluminum capsule. Measurements were performed within an outer lead castle (4 inch thick bricks) and an inner aluminum box (0.25 inch thick) to shield the detector from environmental backgrounds.}
 \label{fig:Yale_exp}
\end{figure}

A photomultiplier tube (PMT) contained within a light-tight housing was used to measure the energy spectrum of the irradiated NaI crystal. We optically coupled the fused silica face of the detector to a Hamamatsu R12669 photomultiplier tube using EJ-550 optical gel from Eljen Technology. The PMT was outfitted with a custom-designed negative-bias base of the same design used in the \mbox{COSINE-100} experiment~\cite{cosine_det}. The entire PMT-sodium iodide detector setup was contained within a small aluminum capsule in order to stabilize the coupling and shield the PMT from external light. All measurements took place within a lead shielding structure that provided at least \SI{10}{\cm} of shielding on all sides of the detector. Figure~\ref{fig:Yale_exp} provides a schematic view of the assembled aluminum capsule and its position within the lead castle. The detector was further contained within a 0.25-inch (\SI{0.64}{\cm}) thick aluminum box that was placed inside of the lead castle.

A voltage bias was provided to the PMT with a CAEN V6533M high voltage module. Due to the limited dynamic range of the PMT-voltage divider combination used, we performed measurements at three different bias voltages: \SI{-1500}, \SI{-1000}, and \SI{-800}{V}. Operation of the PMT at \SI{-1500}{V} results in the best resolution of the energy spectrum in the low-energy region, whereas operation at \SI{-800}{V} allowed access to energies above \SI{1}{MeV}. The PMT signal was recorded by a CAEN V1730 digitizer, featuring a 500 MS/s sampling rate and 14-bit resolution over a 2 V dynamic range. Triggers were generated by the digitizer when a waveform voltage that exceeded a preset threshold was detected. Due to baseline shifts, this trigger threshold varied between different data runs, but was chosen such that $\sim 100\%$ trigger efficiency was maintained within our energy regions of interest. Upon generating a trigger, an \SI{8}{\micro s} long waveform trace was recorded by the digitizer. The recorded waveforms comprise pre-trigger and post-trigger regions, and the digitizer was configured so that the trigger occurred at the \SI{1.6}{\micro s} mark of the recorded waveforms.

After completion of a data run, the recorded waveforms were analyzed offline to extract higher-level physics quantities of interest, primarily the detector's energy spectrum. As an initial step, we computed baseline-subtracted waveforms, which corrected for baseline fluctuations between events. For each event, the baseline was calculated as the average charge of the first \SI{1}{\micro s} within the pre-trigger region of the event waveform. The calculated baseline value was then subtracted from every sample in the recorded waveform. Once the baseline subtraction was performed, we computed the integrated charge of the waveform within a \SI{5}{\micro s} long integration window beginning \SI{1.4}{\micro s} after the start of the event. In addition, we also extracted the maximum charge of the waveform, which was defined as the maximum charge recorded by the digitizer within the full \SI{8}{\micro s} window of the event. After computing these higher-level quantities, we generated the energy spectrum for the processed run.

To convert the integrated charge spectrum to an energy spectrum in units of kiloelectron-volts, removing variations due to the PMT gain, we adopted a calibration function that assumed a proportional response between the integrated charge of the signal produced by the PMT and the amount of energy deposited within the NaI detector. Though this calibration method neglected the known nonlinear nature of the light response of NaI detectors \cite{knoll,rooney_1997,mengesha_1998}, this effect was accounted for in the simulated energy spectra, as detailed in Section~\ref{sec:analysis}. The calibration factor was independently determined for each data run and was defined as the ratio of the mean value of a fitted decay line in the charge integral spectrum to the known energy of the decay. The specific decay line used for calibration depended on the PMT bias voltage of the run. The \SI{-1500}{V} gain setting used the $^{125}$I feature at \SI{40}{keV}, except for the final three measurements due to the decay of this isotope, whereupon the calibration was performed using K x-rays from a $^{133}$Ba source. The \SI{-1000}{V} gain setting was calibrated using the \SI{145}{ keV} internal line from $^{125m}$Te, while the \SI{-800}{V} gain setting was calibrated using an external $^{137}$Cs source.

\subsubsection{Run Summary}
\label{sec:counting:run-sum}

We began measurements of the energy spectra of the irradiated NaI detector in December 2020 and performed subsequent measurements roughly every month until December 2021. Knowledge of the time evolution of the detector's energy spectrum allowed us to distinguish isotopes with decay lines at similar energies but different half-lives. In December 2020 and January 2021, the recorded data runs were two hours in duration and were performed at PMT bias voltages of \SI{-1500} and \SI{-1000}{V}. In February 2021, an eight-hour run at a bias voltage of \SI{-800}{V} was added to the data runs collected each month. Each measurement of the spectrum of the irradiated detector at the \SI{-800}{V} setting was accompanied by a \SI{5}{\minute} calibration run using a \isotope{Cs}{137} source placed \SI{21}{inches} (\SI{53}{\cm}) from the front face of the detector encapsulation, as seen in Figure~\ref{fig:Yale_exp}. This calibration run was necessitated by the lack of suitable calibration peaks within the energy region of interest for the data collected at the \SI{-800}{V} setting. Calibration runs for the final three \SI{-1500}{V} runs were performed with a \isotope{Ba}{133} source at this same position. Beginning with the April 2021 run, runs at the \SI{-1500}{V} and \SI{-1000}{V} settings were extended to eight hours in length, due in part to the decreasing activity of the detector over time. This data-taking configuration continued for all subsequent data sets.

We also measured the energy spectrum of the environmental background radiation using the fourth, unirradiated NaI detector in January 2022. In this measurement, the unirradiated detector was placed in the same position within the lead shield as Crystal A.
\section {Analysis}
\label{sec:analysis}

\subsection{Decay Simulation}
\label{sec:decaysim}
We used a Geant4 (version 10.05.p01) simulation of the crystal measurement setup to model the expected spectral contributions from neutron-activated isotopes in the irradiated NaI crystal.

We simulated the decays of the 75 highest-activity isotopes predicted to have been produced in the crystal (by the beam activation simulation described in Section~\ref{sec:fluence}) with half-lives $>9$ days. The spatial distribution of isotopes was assumed to follow the LANSCE beam profile inside the NaI crystal. Radioisotopes activated in the crystal housing components were also simulated under the simplifying assumption that they were distributed uniformly within the housing materials. The decay and energy deposition processes were modeled using G4EmStandardPhysics$\_$option4, G4RadioactiveDecayPhysics and G4HadronElasticPhysics \cite{allison2016recent}.

While Geant4 can accurately model the energy deposits from radioactive isotopes in the NaI crystal, the measured scintillation light output in the crystal is known to be a non-linear function of the energy deposit \cite{rooney_1997}. Moreover, because a single energy deposit can be partitioned in multiple ways, a microscopic model of the light yield nonlinearity is required, rather than a simple correction to the total energy deposited per event \cite{Moses2008}. In this work, we use the semi-empirical model of light yield nonlinearity developed by Payne \textit{et al.} \cite{payne2011nonproportionality}, expressed as a differential light yield ($LY$) correction factor $\eta$ to the simulated deposited energy ($E$) per step:
\begin{align}
\label{eq:lightyield}
d(LY) = dE \times \eta(dE/dx)
\end{align}
where $\eta$ depends on the stopping power $dE/dx$ of the particle in the NaI crystal
\begin{align}
\eta(dE/dx) = \eta_0\frac{1-\eta_{e/h}\exp\left[\frac{-(dE/dx)}{(dE/dx)_{ONS}}\right]}{1+\left[\frac{(dE/dx)}{(dE/dx)_{BIRKS}}\right]} 
\end{align}
$\eta_0$ is a calibration scaling factor, $(dE/dx)_{ONS} = 36.4$ MeV/cm is a fixed parameter in the light yield model from Ref.~\cite{payne2011nonproportionality}, and we have taken $\eta_{e/h} = 0.33$ and $(dE/dx)_{BIRKS} = 366$ MeV/cm to describe the experimental data. 

At each simulated interaction step of the radioactive decay products, the scintillation light produced is calculated according to Eq.~\ref{eq:lightyield} using the stopping power from Ref.~\cite{payne2011nonproportionality}. To minimise effects due to large changes in particle energy during a single simulation step, the Geant4 production cut and maximum step size were set at \SI{5}{\micro\meter} and \SI{200}{\nano\meter}, respectively. The total scintillation light output of the decay event is then taken to be the sum of the light output at each step. As a cross-check on our methodology, we successfully replicated the light yield non-linearity predicted by a numerical integration of this semi-empirical model with a series of simulated monoenergetic electrons in Geant4.

\begin{figure}
    \centering
    \includegraphics[width=0.45\textwidth]{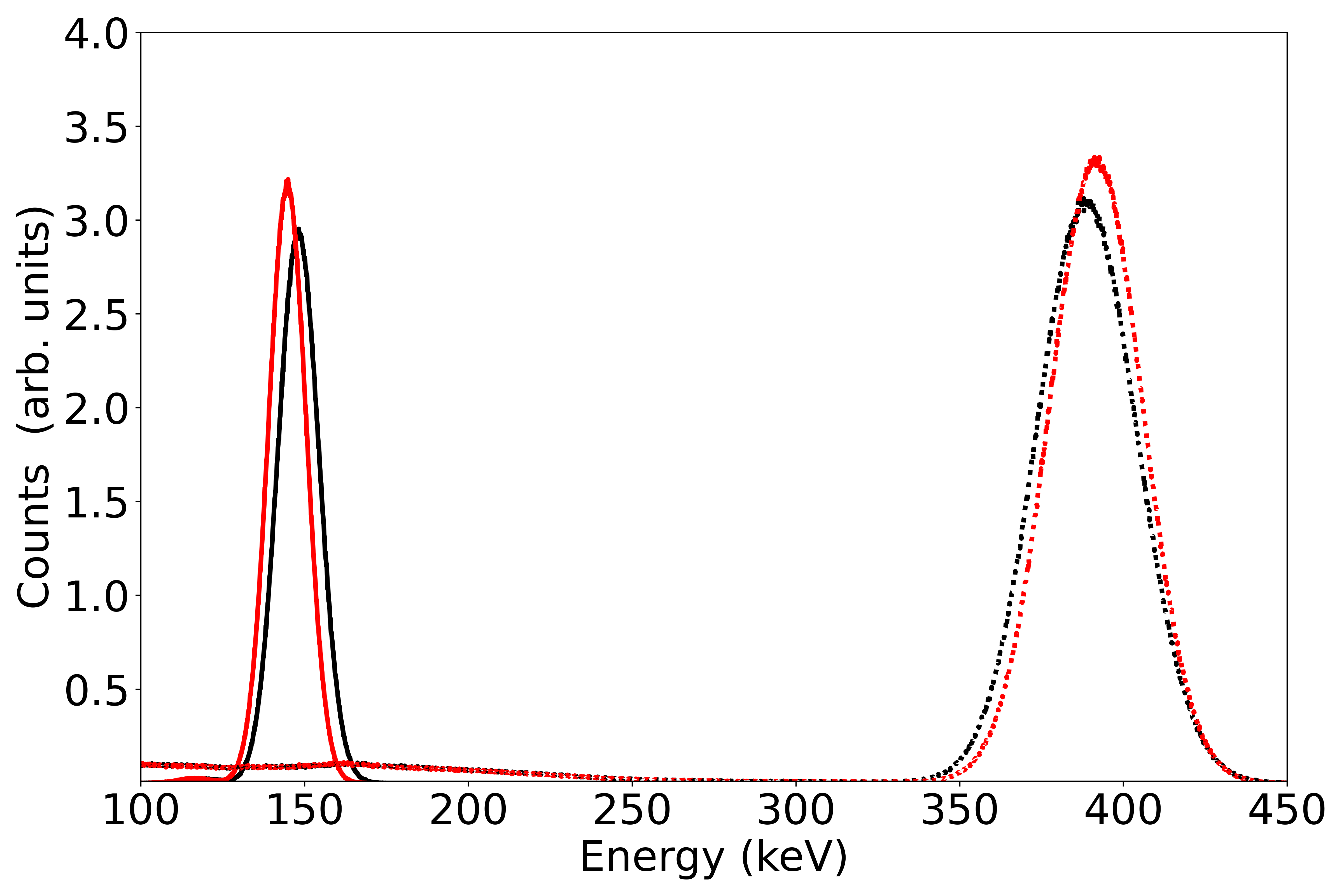}
    \caption{The simulated spectra of uncorrected energy deposits (red) and after applying the light yield non-linearity model (black) in the crystal for $^{125m}$Te (solid lines) and $^{113m}$In (dashed lines). Note that the inclusion of the light yield non-linearity shifts the peaks in opposite directions as described further in the text.}
    \label{fig:LYcompare}
\end{figure}

Figure \ref{fig:LYcompare} compares the spectra obtained from the simulated light yield to that obtained from the energy deposited, for $^{125m}$Te and $^{113m}$In -- both are metastable isotopes which emit fixed amounts of energy per decay. There is an obvious difference between the photopeak locations given by the light yield model versus the energy deposits, which illustrates the effect of accounting for the scintillation non-linearity. The differences are in the opposite sense for $^{125m}$Te compared to $^{113m}$In. We attribute this to the fact that  $^{125m}$Te involves a cascaded transition of two lower energy gamma rays, whereas the $^{113m}$In decay involves a single gamma ray emission. The NaI light yield nonlinearity increases with decreasing energy \cite{payne2011nonproportionality}, $^{125m}$Te leading to a positive shift for that isotope. All simulated spectra used in our fits were corrected for scintillation non-linearity with the exception of $^{3}$H, as discussed in the next section.




\subsection{Fitting}
Our fit model added the binned contributions of the simulated isotope spectra and background components as summarised in Table~\ref{tab:FitSummary}. The isotopes included in the fit do not include all simulated isotopes. Since the measurement campaign described in this paper started 411.12 days after the irradiation, isotopes with $<$20 day half-lives were excluded from the analysis. The exceptions are those isotopes that are fed by longer-lived parents. In those cases, isotopes with $<$20 day half-lives were fixed to the decay rates of their parents. A preliminary analysis of all remaining isotopes indicated that some did not significantly contribute to the observed energy spectrum. A likelihood ratio test was used to remove isotopes from the fit that did not contribute in a statistically significant way. We also included the environmental background in our model, as measured by the unirradiated NaI crystal in the same geometry. An external $^{22}$Na background component from activation of the crystal housing by the neutron beam was also included, taken as the total contribution from $^{22}$Na. All other contributions from the crystal housing were simulated but found to be negligible.


\begin{table}[t]
\centering
\begin{tabular}{c c}
\hline
Component & Notes \\
\hline
\vrule width 0pt height 2.2ex
$^3$H       &   \\
$^{22}$Na   &    \\
$^{109}$Cd  & $^{109m}$Ag assigned equal activity \\
$^{113}$Sn  & $^{113m}$In assigned equal activity \\
$^{121m}$Te & $^{121}$Te assigned equal activity\\
$^{123m}$Te &  \\
$^{125}$I   &  \\
$^{125m}$Te &   \\
$^{127m}$Te &   \\
$^{123}$Sn  & Nuisance parameter \\
$^{101}$Rh  & Nuisance parameter \\
$^{22}$Na in housing   & Nuisance parameter \\
Room background & Nuisance parameter   \\
\hline
\end{tabular}
\caption{A summary of the spectral components included in the fit models (see text for details).}
\label{tab:FitSummary}
\end{table}

In order to convert from the simulated scintillation light output to a nominal energy scale we normalized the simulated spectra to the internal 40.12 keV $^{125}$I, 144.755 keV $^{125m}$Te, and 1275.5 keV $^{22}$Na peaks at high, medium, and low gain settings respectively. The simulated scintillation spectra were also smeared with a Gaussian function so as to reproduce the measured resolution of the NaI crystals.  We used these smeared simulated spectra generated by the light yield model described in Section~\ref{sec:decaysim} without modification for all isotopes except $^3$H, $^{123m}$Te, and $^{121m}$Te. The modifications to the metastable tellurium spectra were limited to their peaks at 247.4 keV and 294.0 keV. These peaks consist of summed contributions of two nuclear transitions where the relative NaI light yield is quite different between the two transitions. Although the light yield model improved the performance of the fit relative to simply using the energy deposited in the crystal, the residual modeling error of these features was large enough to warrant manually scaling the simulated photopeaks by approximately 1\% to the measured values, in order to prevent systematic uncertainties in the model. Using the light yield model for the $^3$H spectrum gave a measurably worse fit to the low-energy part of the spectrum, both by eye and in terms of the fit likelihood. We do not know why the $^3$H performed poorly, while all other spectral components were improved by the light yield model. We speculate that our calibration of the Payne model \cite{payne2011nonproportionality} with gamma rays may be partially to blame, since $^3$H is a pure beta emitter. Determining the cause of this modeling deficiency is beyond the scope of this work. We have used the $^3$H energy deposit in our analysis, since that best reproduces the measured spectrum. We have also taken this difference in modeling approach for $^3$H as a systematic uncertainty.

The fits maximised the joint likelihood of the model, given the measured low, medium, and high gain data. The fit ranges at these gain settings were \SIrange{2.55}{50}{keV}, \SIrange{50}{450}{keV}, and \SIrange{450}{1600}{keV}, respectively. The upper limits to the fit ranges were set conservatively to ensure that there was negligible influence of photomultiplier nonlinearity in the data. The regions of nonlinearity were determined by examining both the distribution of the waveform height to charge integral ratio and comparisons of the calibrated energy spectra with simulated energy spectra. The 2.55 keV minimum energy was driven by a trigger threshold analysis of the experimental measurements to ensure our measurements had $\sim 100\%$ trigger efficiency. The fit was achieved using a Markov Chain Monte Carlo (MCMC) sampling of the fit parameters, which made use of the Metropolis-Hastings algorithm \cite{Robert2005}. The MCMC sampling typically achieved a better maximum likelihood than a gradient descent optimiser, presumably due to the presence of local minima. The MCMC sampling was continued for long enough to sample the posterior distributions of the fit parameters, allowing a reasonable estimate of the fitting uncertainty.

The fitting methodology proceeded in two steps. First, a month-by-month analysis treated each set of measurements at a given time independently of measurements at other times. The monthly fits allowed a comparison of the time dependence of the fitted spectral components with the expected behaviour given by the decay rate of the isotopes. This analysis was used as a cross-check to ensure that we correctly identified the features of the spectrum with the corresponding isotopes. Second, a simultaneous fitting analysis was performed to extract the final isotopic activities. The simultaneous fit considered all spectrum measurements as part of a single combined likelihood fit, where the time dependence of the isotopic spectral components was fixed by the known nuclear decay half-lives for the relevant isotopes. The details and results of the two fitting steps are described in detail below. 

\begin{figure}[ht]
\centering     
\subfigure[ ]{
\label{fig:Feb_a}\includegraphics[width=0.45\textwidth]{
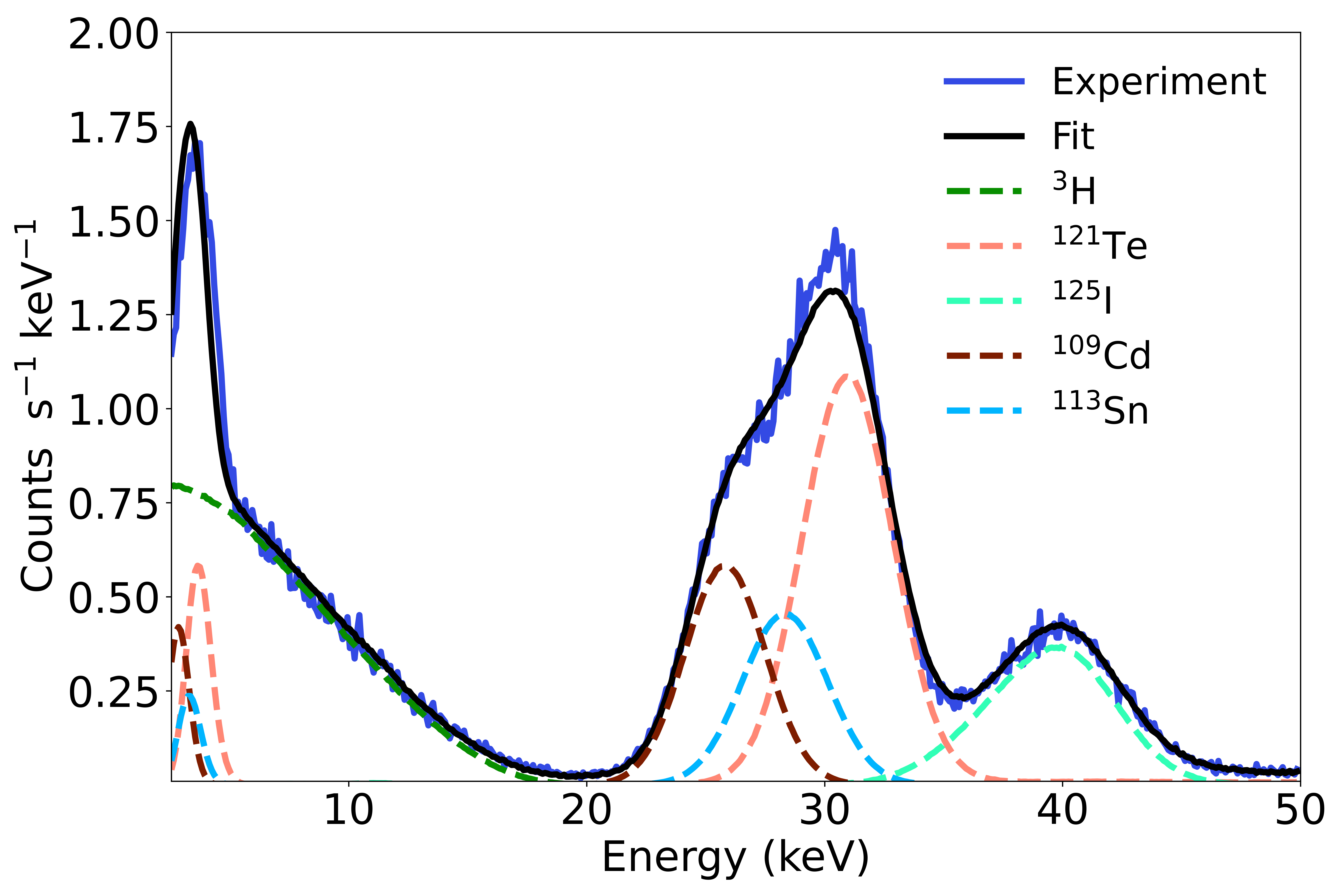}}
\subfigure[ ]{
\label{fig:Feb_b}\includegraphics[width=0.45\textwidth]{
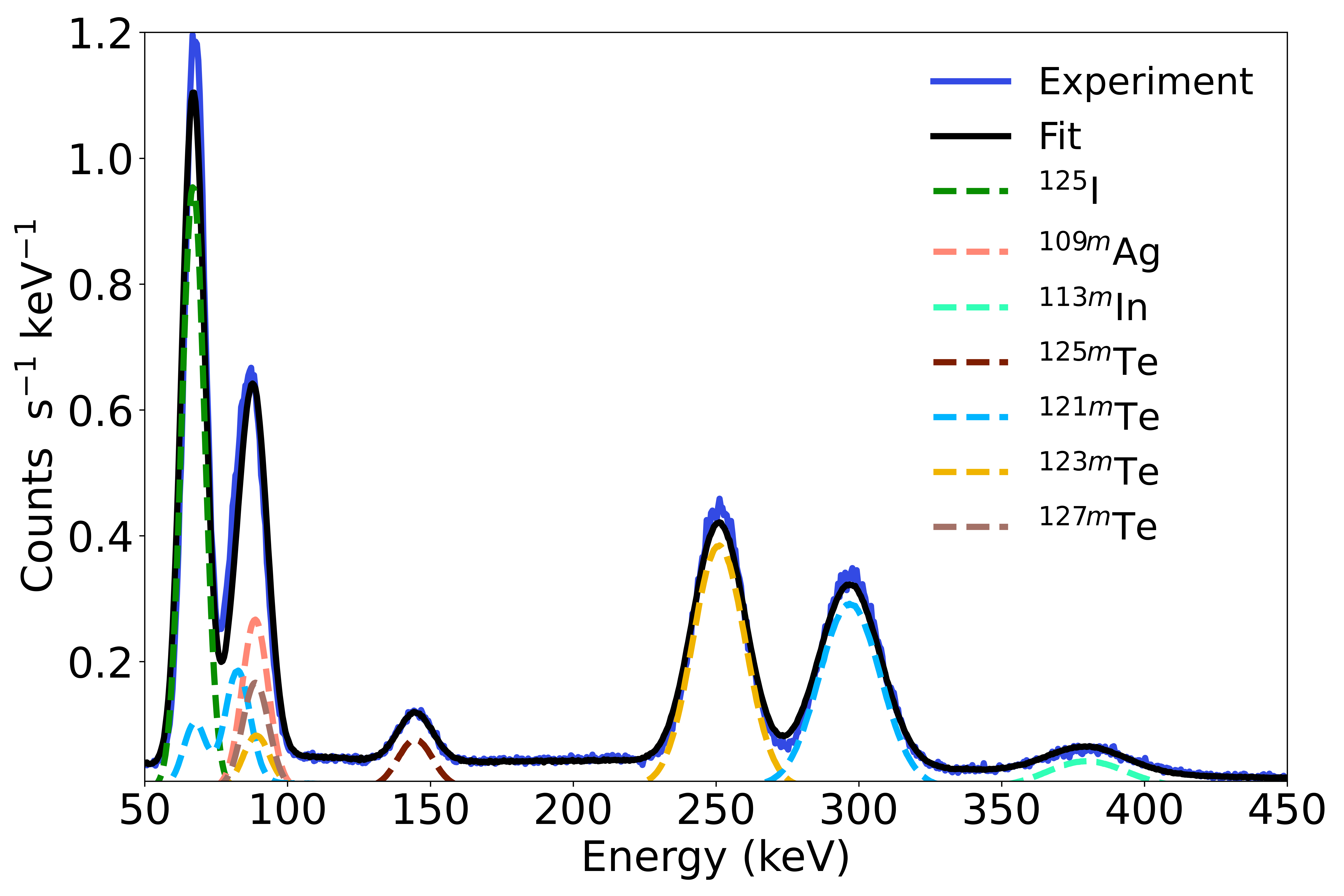}}
\subfigure[ ]{
\label{fig:Feb_c}\includegraphics[width=0.45\textwidth]{
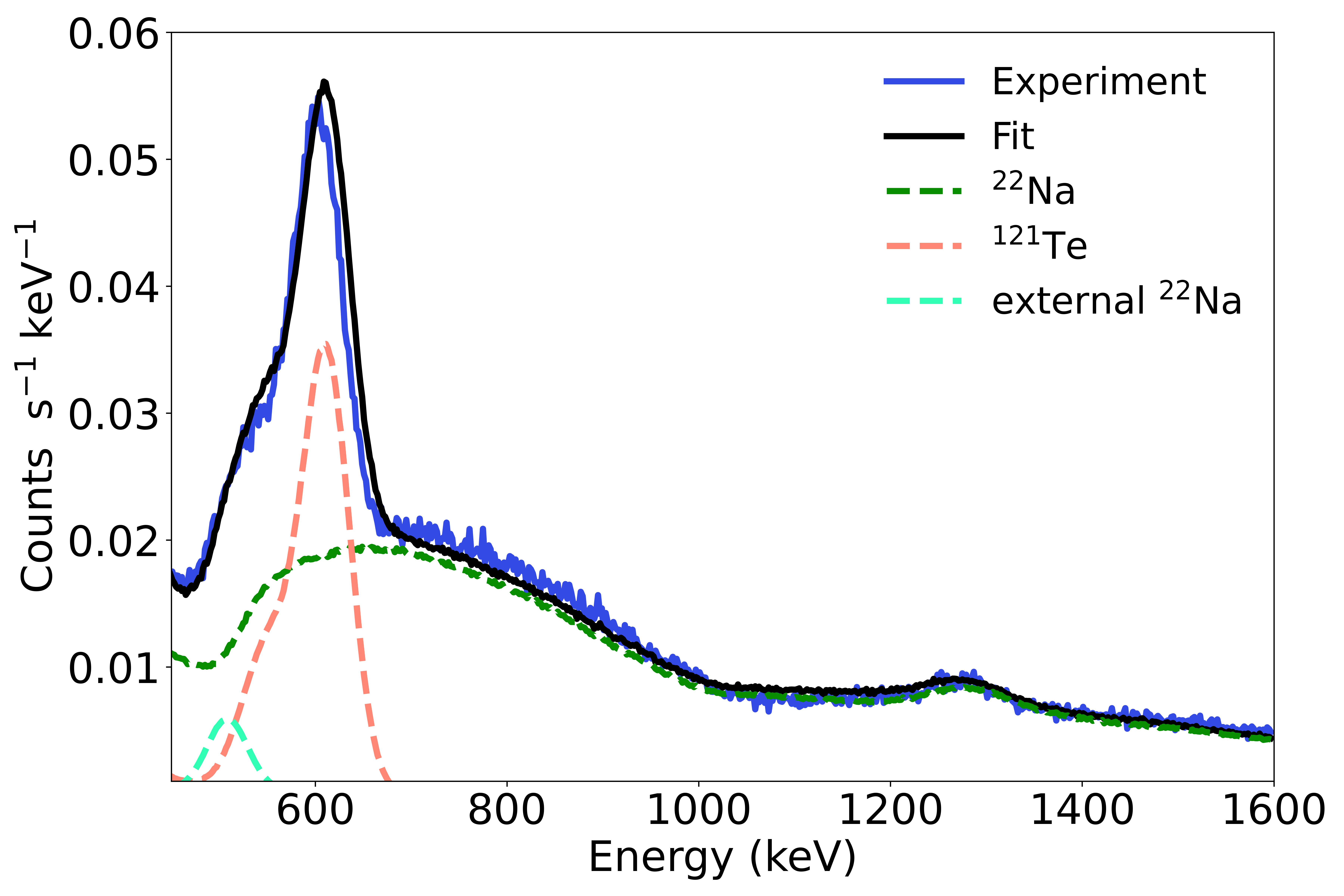}}
\caption{The experimental measurement of the induced activity at February 2021 (blue) and the result of the month-by-month fitting (black) at PMT voltages of (a) -1500 V, (b) -1000 V, and (c) -800 V. The most important isotopes contributing to the fit in each energy range are also shown (dashed lines).}
\label{fig:FebFits}
\end{figure}

\subsubsection{Month-by-month fitting}
An example of a measurement at the three gain settings and the maximum likelihood fit estimate is shown in Figure~\ref{fig:FebFits}. The spectrum exhibits a number of readily identifiable components, including the $^{3}$H beta distribution at low-energy, multiple x-ray and gamma ray peaks, and a continuum contribution from $^{22}$Na that dominates at high-energy. Indeed, most isotopic components create at least one feature in the spectrum where they are the sole or dominant contributor, and this has allowed the fit to constrain their activity at each measurement in what is a relatively complicated spectrum. Exceptions to this are the $^{127m}$Te/$^{127}$Te chain, which has its largest contribution as a minority component of the unresolved peak at approximately 90 keV, as well as the components designated as nuisance parameters that are discussed below.

\begin{figure}
    \centering
    \includegraphics[width=0.45\textwidth]{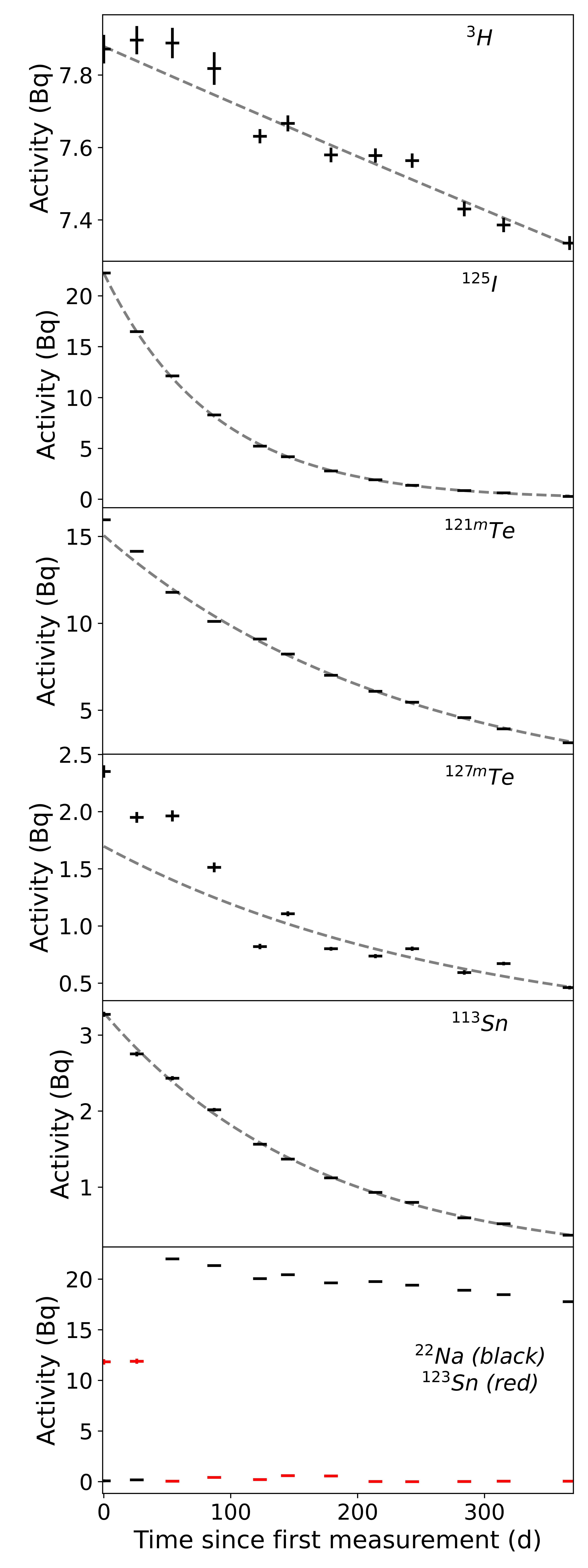}
    \caption{The time dependent activity returned by the month-by-month fitting. The dashed lines indicate the best fits to an exponential decay. The deduced halflives are compared with literature values in Figure~\ref{fig:HLfits}}
    \label{fig:MonthlyDecay}
\end{figure}

The time-dependence of activities for selected isotopes, as determined by the month-by-month fits, are plotted in Figure~\ref{fig:MonthlyDecay}. Most isotopic components appear to follow an exponential decay. Over the first two months the $^{22}$Na and $^{123}$Sn do not follow an exponential decay, with the activities anti-correlated. This anomalous behaviour appears to be due to the lack of data in the high energy range for the first two months, which would otherwise constrain the $^{22}$Na component. At lower energies, both isotopes contribute a sub-dominant broad continuous spectrum, suggesting the fit finds it difficult to discriminate between the relative contributions. Other isotopic contributions appear to be less affected, although there are similar anomalies for the external $^{22}$Na and room background components.

\begin{figure}
    \centering
    \includegraphics[width=\columnwidth]{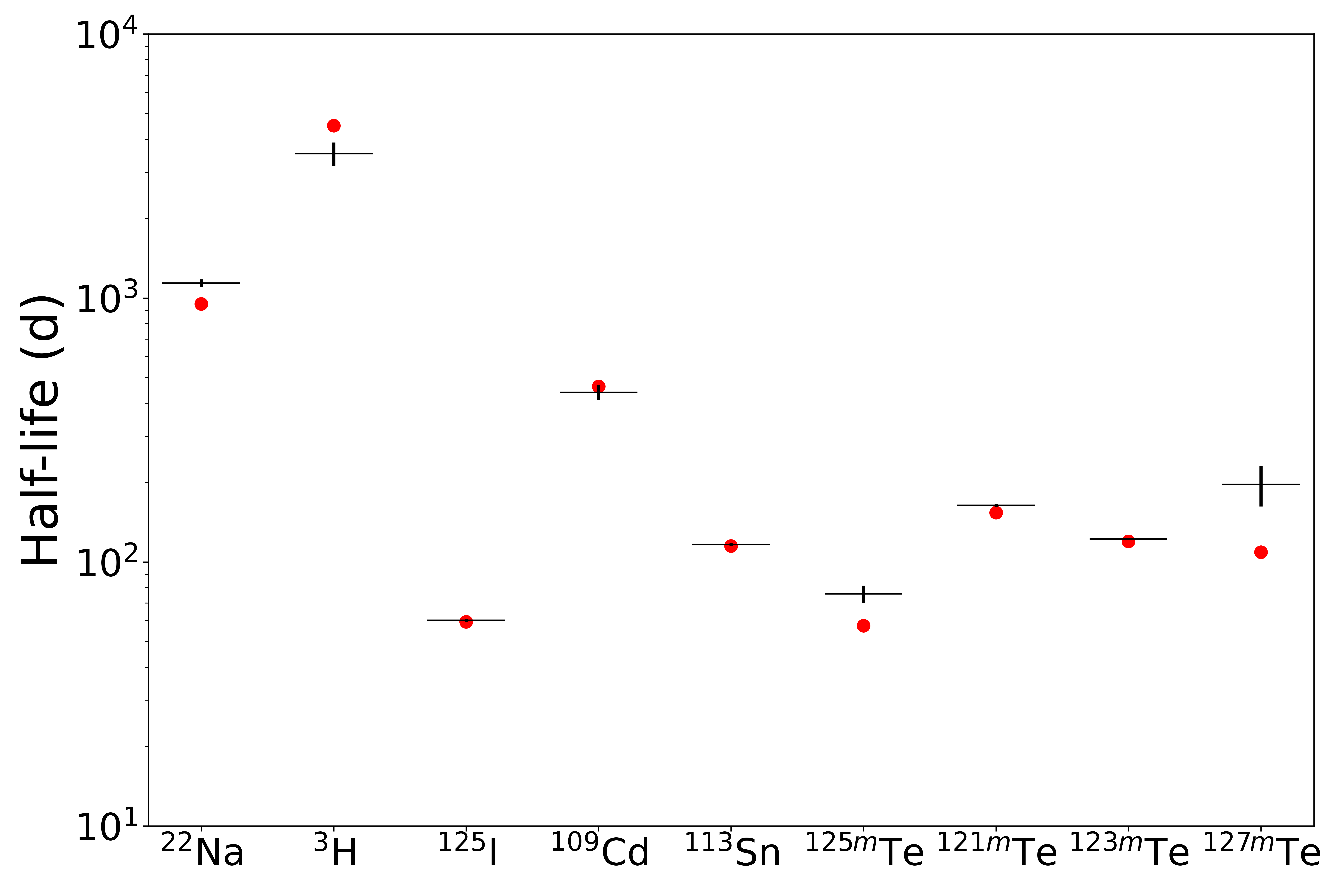}
    \caption{The half-lives for the isotopes produced in the NaI, as determined by an exponential fit to their monthly activities returned by the fit (black), along with the fit uncertainties. The nuclear data values are also shown (red).}
    \label{fig:HLfits}
\end{figure}

We have fitted the time-dependent behaviour of the isotopes with an exponential function for all but the first two months (see Figure~\ref{fig:HLfits}), as a check of the fit model's robustness and a cross check that the spectral features are being associated with the correct isotope. We note that even after excluding the first two months, $^{123}$Sn and $^{101}$Rh do not appear to follow an exponential decay. These isotopes are only present in the fit at sub-dominant levels and we have taken these, along with the room background and external $^{22}$Na component as nuisance parameters in our model. The fitted half-lives for the remaining isotopes agree fairly well with the nuclear data, with the ratio of the fitted value to the accepted nuclear data half-life deviating most from unity for $^{127m}$Te (1.80 $\pm$ 0.32), followed by $^{125m}$Te (1.32 $\pm$ 0.10). Overall, the results in Figure~\ref{fig:HLfits} are sufficiently close to the expected nuclear data half-lives to confirm that our model is correctly identifying the isotopes associated with the measured spectral components.

\subsubsection{Simultaneous fitting} 
Once we confirmed that we had correctly identified the spectral components, for our final results we analyzed the data by considering all measurements, both over time and at different voltages, as part of a single combined likelihood fit. The external background was taken to be constant in time, while the time dependence of the activated isotopic spectral components was fixed by the relevant decay rate given by the nuclear data. For all activation components except $^{125m}$Te, the time profile of the activity over the span of the measurements is expected to be a single exponential, dominated by the half-life of a single isotope (see details in Section~\ref{sec:production_rates}). $^{125m}$Te (57.4 day half-life) can either be directly produced by the beam or be produced through the decay of $^{125}$Sb (2.76 year half-life). Both halflives are long enough that both direct production and radioactive decay could contribute significantly to the $^{125m}$Te decays seen in the data. Furthermore, since $^{125}$Sb did not contribute in a statistically meaningful way to the fit, its contribution is only visible to our model via the decay behaviour of $^{125m}$Te. It is impossible to determine the time dependence of $^{125m}$Te without imposing a production model predicting the ratio of direct activation to feeding, so we have allowed its monthly contributions to float in the fit and have recovered an estimate of the $^{125}$Sb activity using a fit to that time dependence.

\begin{figure*}
    \centering
    \includegraphics[width=\textwidth]{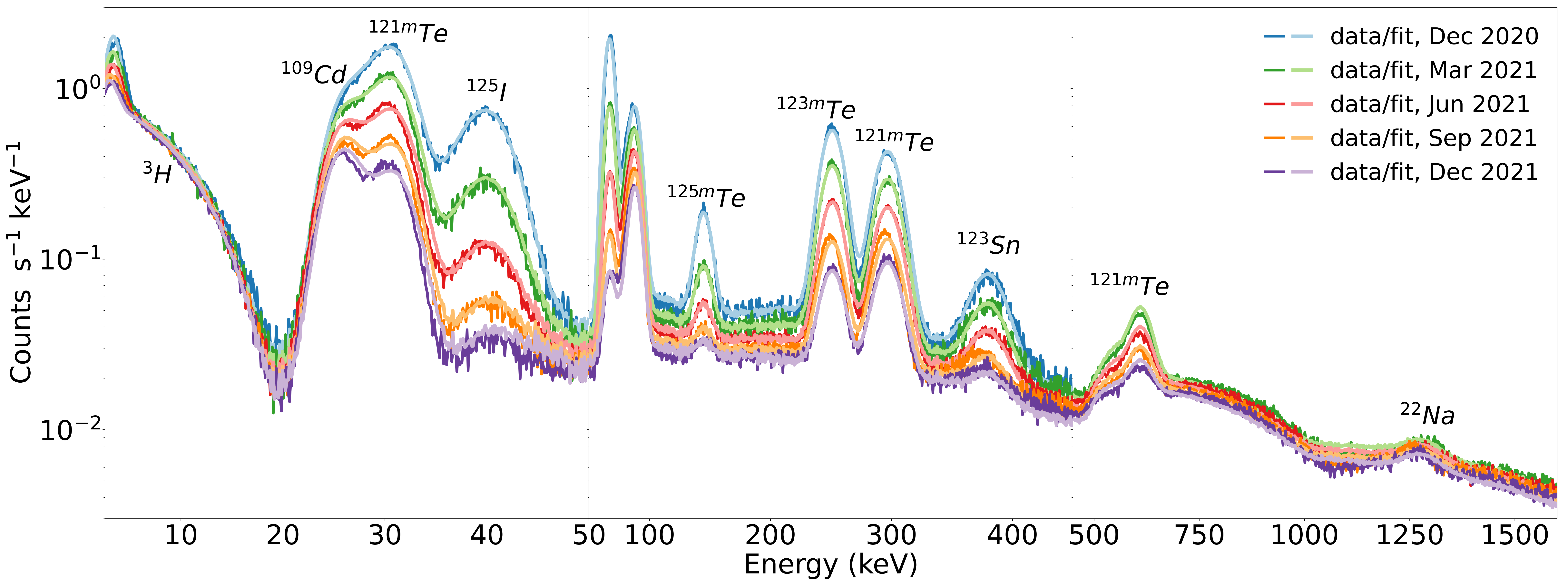}
    \caption{The data across the entire energy range for the December 2020 (blue), March 2021 (green), Jun 2021 (red), September 2021 (orange), and December 2021 (purple) measurements, and the simultaneous fit at those times with a corresponding colour and a lighter shade. The spectral features are labelled with the identity of the isotope primarily responsible for each feature.}
    \label{fig:SimultaneousSpectra}
\end{figure*}

The maximum likelihood fit to the data at four different times, separated by approximately 3 months each, is shown in Figure~\ref{fig:SimultaneousSpectra}. These measurements illustrate the decay of the various spectral components and their reproduction by the fit model. The overall likelihood for the simultaneous fitting model was less than the combined likelihood for the monthly fitting models, and a likelihood ratio test suggested that there was a non-statistical component to this decrease. We attribute the difference to deficiencies in our energy scale and energy resolution model, which are accounted for in the systematic uncertainty estimation detailed below.

Excluding the nuisance parameters, the statistical uncertainties in the isotopic activities estimated using the MCMC sampling was less than 1\%, with the lowest uncertainty of 0.07\% coming from $^{22}$Na, which is well-constrained over a large swathe of the high-energy spectrum. Of the isotopes that were fit with a known exponential decay (all except $^{125m}$Te), the $^{127m}$Te/$^{127}$Te activity had the highest statistical uncertainty at 0.73 \%, due to its sub-dominant contribution to the spectrum. 

\begin{figure}
    \centering
    \includegraphics[width=0.47\textwidth]{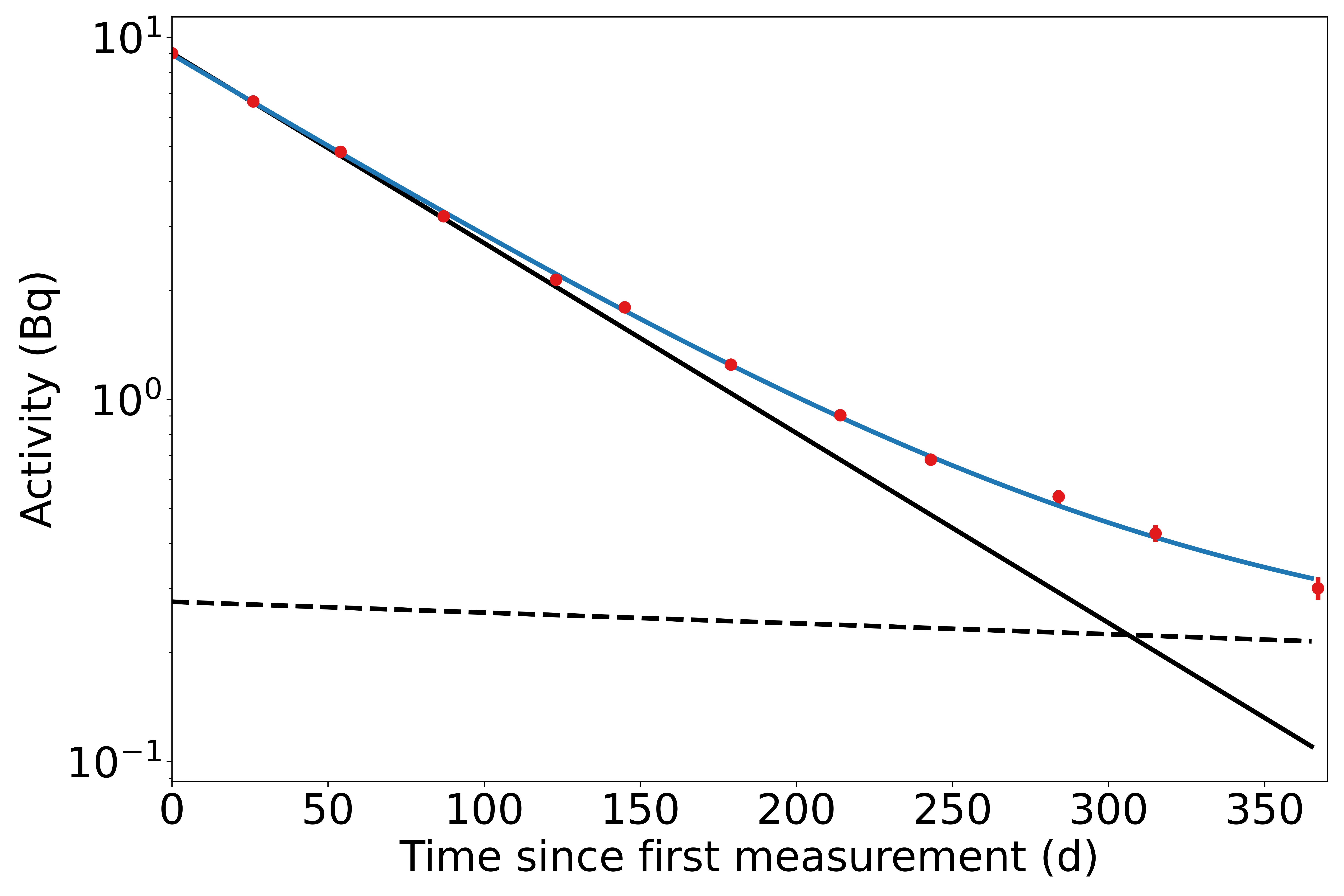}
    \caption{The time dependence of the $^{125m}$Te activity. The data (red) is well-described using a model (blue) with decays from initial $^{125m}$Te production as well as feeding from $^{125}$Sb. The individual exponential components due to $^{125m}$Te (solid black) and $^{125}$Sb (dashed black) are shown for comparison.}
    \label{fig:Te125mDecay}
\end{figure}

The time-dependent activity of $^{125m}$Te is less well-constrained because no \textit{a priori} time dependence was assumed by the simultaneous fit. Its activity over time is shown in Figure~\ref{fig:Te125mDecay} where it can be seen that a single exponential fit is inadequate to describe the data. Instead we have used a double exponential fit, with decay rates constrained to those known for $^{125m}$Te and $^{125}$Sb, to extract the activities of the two components: direct beam-produced $^{125m}$Te, and $^{125m}$Te produced through the decay of the longer-lived $^{125}$Sb. We use the fitted exponential components to extract the rate of direct beam-produced $^{125m}$Te (\SI{8.68 \pm 0.95}{\becquerel}) and decay-produced $^{125m}$Te (\SI{0.28 \pm 0.03}{\becquerel}) at the reference time. The decay-produced $^{125m}$Te rate is then used to calculate the $^{125}$Sb rate using the known decay branching ratio.   

\begin{table}[t]
\centering
\begin{tabular}{c c c c}
\hline
Isotope & Activity \\
 & [Bq] \\
\hline
\vrule width 0pt height 2.2ex
$^3$H       & 7.797 $\pm$ 0.207 (sys) $\pm$ 0.007 (stat)    \\
$^{22}$Na   & 22.987 $\pm$ 0.971 (sys) $\pm$ 0.016 (stat)   \\
$^{109}$Cd  & 3.534 $\pm$ 0.409 (sys) $\pm$ 0.005 (stat)    \\
$^{113}$Sn  & 3.231 $\pm$ 0.235 (sys) $\pm$ 0.010 (stat)    \\
$^{121m}$Te & 15.834 $\pm$ 0.574 (sys) $\pm$ 0.012 (stat)   \\
$^{123m}$Te & 13.791 $\pm$ 0.693 (sys) $\pm$ 0.015 (stat)   \\
$^{125}$I   & 22.361 $\pm$ 1.359 (sys) $\pm$ 0.027 (stat)   \\
$^{125}$Sb  & 1.206 $\pm$ 0.132 (sys) $\pm$ 0.069 (stat)    \\
$^{125m}$Te & 8.683 $\pm$ 0.948 (sys) $\pm$ 0.077 (stat)    \\
$^{127m}$Te & 2.321 $\pm$ 1.763 (sys) $\pm$ 0.017 (stat)    \\
\hline
\end{tabular}
\caption{The measured activities at the reference time of 411.12 days after the irradiation, for isotopes included in the simultaneous fitting model. The $^{125m}$Te and $^{125}$Sb activities were inferred from the time-dependence of the $^{125m}$Te peak (see text for details).}
\label{tab:CrystalActivity}
\end{table}

The activity values and their uncertainties are reported in Table~\ref{tab:CrystalActivity}. The $^{125m}$Te activity was fitted using the double exponential described above, and the value reported in the table is the fitted value at the reference time. We considered systematic uncertainties associated with an imperfect knowledge of the energy calibration and energy resolution, as well as due to the treatment of the $^{3}$H energy scale. To estimate the energy calibration uncertainty, we have evaluated the variation of the ratio of the calibration peak used to set the energy scale to a nearby calibration peak at each gain setting, for every measurement. A ratio was preferred to the absolute value to normalise monthly gain variations, which were already accounted for in the calibration process. For the resolution, the variation of the absolute value of the resolution was used. For both the resolution and the energy scale, we have taken the maximum observed variation, independently applied increases and decreases to the energy scale or resolution of the simulated data by this amount, and used these modified simulations to repeat the fitting analysis, returning activity values subject to these variations. We have then taken the magnitude of the largest of these activity variations as an estimate of the standard deviation associated with this systematic. This is a conservative approach to this systematic, as it assumes any variation is both extreme and occurs in the same sense for all gain settings and every measurement time. For the $^{3}$H energy scale modeling, we have taken the difference in activity returned by the fit between the light yield model and energy deposit model for $^{3}$H as an estimate of the standard deviation of this systematic. $^{3}$H was the only isotope appreciably affected by this systematic. The energy calibration, energy resolution, and $^{3}$H modeling systematic uncertainties were combined assuming that they are independent and normally distributed. The systematic uncertainty in the energy scale dominates the overall uncertainty for most isotopes.


\section{Predicted Beam Produced Activities}
\label{sec:production_rates}

\begin{table*}[t]
\begin{center}\setcellgapes{2pt}\makegapedcells \renewcommand\theadfont{\normalsize\bfseries}
\begin{tabular}{cccccccc}
\hline
\hline
Primary  & Feed & \multicolumn{2}{c}{INCLXX}& \multicolumn{2}{c}{BERTINI}& \multicolumn{2}{c}{BIC}\\
Isotope & Isotope & Beam Prod. & Pred. Act. &  Beam Prod. & Pred. Act. & Beam Prod. & Pred. Act. \\
& & [atoms] & [Bq] & [atoms] & [Bq] & [atoms] & [Bq]\\
\hline
\multirow{ 1 }{*}{ $^3$H }& $^3$H & \num{6.05 \pm 0.87 E9} & \num{ 10.1 \pm 1.4 } &  \num{1.56 \pm 0.23 E9} & \num{ 2.62 \pm 0.39 } &  \num{3.89 \pm 0.50 E9} & \num{ 6.50 \pm 0.84 } \\
\hline
\multirow{ 1 }{*}{ $^{22}$Na }& $^{22}$Na & \num{9.05 \pm 0.66 E9} & \num{ 56.6 \pm 4.1 } &  \num{3.54 \pm 0.33 E9} & \num{ 22.1 \pm 2.1 } &  \num{9.59 \pm 0.71 E9} & \num{ 60.0 \pm 4.4 } \\
\hline
\multirow{ 6 }{*}{ \shortstack{$^{109}$Cd, \\$^{109m}$Ag} } & $^{109}$Sb & \num{1.3 \pm 1.2 E5} & $< 0.005$ &  \num{1.58 \pm 0.57 E6} & \num{ 0.01 \pm 0.01 } &  \num{2.2 \pm 1.7 E5} & $< 0.005$ \\
& $^{109}$Sn & \num{3.74 \pm 0.91 E7} & \num{ 0.35 \pm 0.09 } &  \num{1.90 \pm 0.46 E8} & \num{ 1.80 \pm 0.44 } &  \num{3.33 \pm 0.85 E7} & \num{ 0.31 \pm 0.08 } \\
& $^{109}$In & \num{1.57 \pm 0.35 E8} & \num{ 1.48 \pm 0.33 } &  \num{1.10 \pm 0.28 E8} & \num{ 1.04 \pm 0.26 } &  \num{1.30 \pm 0.31 E8} & \num{ 1.23 \pm 0.29 } \\
& $^{109}$Cd & \num{2.56 \pm 0.54 E8} & \num{ 2.44 \pm 0.52 } &  \num{6.2 \pm 1.6 E7} & \num{ 0.59 \pm 0.15 } &  \num{2.11 \pm 0.46 E8} & \num{ 2.01 \pm 0.44 } \\
& \textbf{Sum} & & \textbf{\num{ 4.28 \pm 0.93 }} &  & \textbf{\num{ 3.44 \pm 0.85 }} &  & \textbf{\num{ 3.56 \pm 0.80 }} \\
\hline
\multirow{ 5 }{*}{ \shortstack{$^{113}$Sn, \\$^{113m}$In} } & $^{113}$Te & \num{1.8 \pm 1.1 E5} & $<0.005$ &  \num{1.28 \pm 0.25 E8} & \num{ 0.75 \pm 0.15 } &  \num{1.00 \pm 0.38 E6} & $<0.01$ \\
& $^{113}$Sb & \num{9.8 \pm 1.8 E7} & \num{ 0.56 \pm 0.10 } &  \num{2.05 \pm 0.39 E8} & \num{ 1.18 \pm 0.23 } &  \num{1.21 \pm 0.25 E8} & \num{ 0.70 \pm 0.14 } \\
& $^{113m}$Sn & \num{6.1 \pm 1.0 E8} & \num{ 3.22 \pm 0.56 } &  \num{0.1 \pm 1.0 E5} & $<0.005$ &  \num{7.4 \pm 1.3 E8} & \num{ 3.91 \pm 0.69 } \\
& $^{113}$Sn & \num{1.17 \pm 0.20 E8} & \num{ 0.69 \pm 0.12 } &  \num{4.83 \pm 0.91 E8} & \num{ 2.84 \pm 0.54 } &  \num{1.39 \pm 0.24 E8} & \num{ 0.82 \pm 0.14 } \\
& \textbf{Sum} & & \textbf{\num{ 4.47 \pm 0.78 }} &  & \textbf{\num{ 4.77 \pm 0.91 }} &  & \textbf{\num{ 5.43 \pm 0.96 }} \\
\hline
\multirow{ 5 }{*}{ \shortstack{$^{121m}$Te, \\$^{121}$Te} } & $^{121}$Xe & \num{8.9 \pm 2.5 E6} & $<0.005$ &  \num{2.09 \pm 0.57 E7} & $<0.005$ &  \num{1.61 \pm 0.47 E7} & $<0.005$ \\
& $^{121}$I & \num{1.24 \pm 0.11 E9} & \num{ 0.02 \pm 0.01 } &  \num{5.10 \pm 0.42 E9} & \num{ 0.06 \pm 0.05 } &  \num{2.00 \pm 0.18 E9} & \num{ 0.03 \pm 0.02 } \\
& $^{121m}$Te & \num{1.87 \pm 0.18 E9} & \num{ 16.5 \pm 1.6 } &  \num{7.2 \pm 3.0 E5} & $<0.01$ &  \num{2.71 \pm 0.25 E9} & \num{ 23.98 \pm 2.28 } \\
& $^{121}$Te & \num{3.95 \pm 0.37 E8} & $<0.005$ &  \num{2.72 \pm 0.25 E9} & $<0.005$ &  \num{5.45 \pm 0.49 E8} & $<0.005$ \\
& \textbf{Sum} & & \textbf{\num{ 16.6 \pm 1.6 }} &  & \textbf{\num{ 0.07 \pm 0.05 }} &  & \textbf{\num{ 24.0 \pm 2.3 }} \\
\hline
\multirow{ 1 }{*}{ $^{123m}$Te }& $^{123m}$Te & \num{2.17 \pm 0.18 E9} & \num{ 13.4 \pm 1.1 } &  \num{6.4 \pm 2.6 E5} & $<0.005$ &  \num{2.47 \pm 0.20 E9} & \num{ 15.2 \pm 1.2 } \\
\hline
\multirow{ 1 }{*}{ $^{125}$I }& $^{125}$I & \num{1.18 \pm 0.08 E10} & \num{ 13.10 \pm 0.90 } &  \num{2.18 \pm 0.15 E10} & \num{ 24.4 \pm 1.6 } &  \num{1.61 \pm 0.11 E10} & \num{ 18.0 \pm 1.2 } \\
\hline
\multirow{ 4 }{*}{ $^{125}$Sb } & $^{125}$Sn & \num{3.06 \pm 0.42 E7} & \num{ 0.18 \pm 0.02 } &  \num{9.0 \pm 1.9 E6} & \num{ 0.05 \pm 0.01 } &  \num{7.9 \pm 2.0 E6} & \num{ 0.05 \pm 0.01 } \\
& $^{125}$Sb & \num{5.80 \pm 0.54 E8} & \num{ 3.62 \pm 0.36 } &  \num{1.46 \pm 0.17 E8} & \num{ 0.91 \pm 0.11 } &  \num{2.04 \pm 0.22 E8} & \num{ 1.27 \pm 0.15 } \\
& \textbf{Sum} & & \textbf{\num{ 3.80 \pm 0.38 }} &  & \textbf{\num{ 0.96 \pm 0.12 }} &  & \textbf{\num{ 1.32 \pm 0.16 }} \\
\hline
\multirow{ 1 }{*}{ $^{125m}$Te } & $^{125m}$Te & \num{1.27 \pm 0.10 E9} & \num{ 1.23 \pm 0.14 } & - & - &  \num{9.59 \pm 0.74 E8} & \num{ 0.93 \pm 0.10 } \\
\hline
\multirow{ 2 }{*}{ \shortstack{$^{127m}$Te, \\$^{127}$Te} }& $^{127m}$Te & \num{1.10 \pm 0.07 E9} & \num{ 5.68 \pm 0.36 } &  \num{1.1 \pm 1.1 E5} & $<0.005$ &  \num{3.46 \pm 0.23 E8} & \num{ 1.79 \pm 0.12 } \\
& & & & & & &\\
\hline
\hline
\end{tabular}
\end{center}
\caption{Predicted production and activity of isotopes from beam activation of NaI crystal C according to Geant4 cross-section models INCLXX, BERTINI, and BIC. The first column indicates the primary isotope(s) that contribute to the visible energy spectrum. The second column lists all the isotopes directly produced by the beam that feed the production of the primary isotope. Under each cross-section model we list the predicted number of feed isotope atoms produced by the beam exposure and the predicted activity of the primary isotope(s) due only to that specific feed isotope at the reference measurement time. The total activity of the primary isotope at the reference time is then given by the sum of the predicted activities across all feed isotopes.
\label{tab:pred_act}}
\end{table*}

\begin{table}[t]
\begin{center}
\begin{tabular}{cccc}
\hline
\hline
Primary  & \multicolumn{3}{c}{Scale Factor}\\
Isotope & INCLXX & BERTINI & BIC\\
\hline
\vrule width 0pt height 2.2ex
$^3$H & \num{ 0.77 \pm 0.11 } & \num{ 2.98 \pm 0.45 } & \num{ 1.20 \pm 0.16 } \\
$^{22}$Na & \num{ 0.41 \pm 0.03 } & \num{ 1.04 \pm 0.11 } &  \num{ 0.38 \pm 0.03 } \\
$^{109}$Cd, $^{109m}$Ag & \num{ 0.83 \pm 0.20 } & \num{ 1.03 \pm 0.28 } & \num{ 0.99 \pm 0.25 } \\
$^{113}$Sn, $^{113m}$In  & \num{ 0.72 \pm 0.14 } & \num{ 0.68 \pm 0.14 } & \num{ 0.60 \pm 0.11 } \\
$^{121m}$Te, $^{121}$Te & \num{ 0.96 \pm 0.10 } & - & \num{ 0.66 \pm 0.07 } \\
$^{123m}$Te & \num{ 1.03 \pm 0.10 } & - & \num{ 0.91 \pm 0.09 } \\
$^{125}$I & \num{ 1.71 \pm 0.16 } & \num{ 0.92 \pm 0.08 } & \num{ 1.24 \pm 0.11 } \\
$^{125}$Sb & \num{ 0.32 \pm 0.05 } & \num{ 1.26 \pm 0.22 } & \num{ 0.92 \pm 0.16 } \\
$^{125m}$Te & \num{ 7.1 \pm 1.1 }  & -  & \num{ 9.3 \pm 1.5 } \\
$^{127m}$Te & \num{ 0.41 \pm 0.31 } & - & \num{ 1.30 \pm 0.99 } \\
\hline
\hline
\end{tabular}
\end{center}
\caption{Scale factor of the measured activity compared to the total predicted activity from each model (see Equation~\ref{eq:scale}).
\label{tab:scale_fact}}
\end{table}
If the neutron beam had an energy spectrum identical to that of cosmic-ray neutrons,  we could simply estimate the cosmogenic production rate by scaling the measured activity by the ratio of the cosmic-ray neutron flux to that of the neutron beam flux. However the beam spectrum falls off faster at higher energies than that of cosmic rays (see Fig.~\ref{fig:fluence}). Thus we must rely on a model for the production cross sections to extrapolate from the beam measurement to the cosmogenic production rate. 

\subsection{Beam Production Predictions}
We evaluated the production of isotopes in the NaI crystal using the Geant4 beam simulation described in Section~\ref{sec:exposure} with three different built-in high energy physics libraries, INCLXX \cite{boudard2013new,mancusi2014extension}, BERTINI \cite{bertini1963low, guthrie1968calculation, bertini1969intranuclear, bertini1971news}, and Binary Cascades (BIC) \cite{folger2004binary}. To evaluate the systematic uncertainties in these predictions we propagated the uncertainties in the beam fluence and target thickness using the thin target formula for the predicted number of atoms of isotope $N_{m,i}$ [atoms], produced by the beam
\begin{linenomath*}
\begin{align}
\label{eq:beam_act}
N_{m, i} = \sum_{x=Na, I}{\rho_{a,x} \int S(E) \cdot \sigma_{m, i,x}(E)~dE}
\end{align}
\end{linenomath*}
where $x$ is either \isotope{Na}{23} or \isotope{I}{127}, $\rho_a$ is the areal number density of the target atoms [\si{\atoms\per \cm\squared}], $S(E)$ is the energy spectrum of neutrons [\si{\neutrons \per \MeV}] and $\sigma_{m, i}(E)$ [cm$^2$] is the isotope production cross section for a given physics list model $m$ and isotope $i$. In addition to being produced directly by neutron interactions from the beam, some isotopes can also be produced through the radioactive decay of other radioisotopes produced by the beam (e.g.\ \isotope{Sn}{125} $\to$ \isotope{Sb}{125} $\to$ \isotope{Te}{125m}). For each isotope that contributes to the fit of the experimentally measured spectrum, we reviewed the corresponding nuclear data sheets and made sure to track the production of all possible parent isotopes in the decay chain that ``feed'' the ``primary'' isotope of interest for the spectral fit. The predicted beam production and uncertainties for all feed isotopes relevant to primary isotopes in the spectral fit of the data are shown in Table~\ref{tab:pred_act}.
\subsection{Decay Corrections}
In order to compare with the experimental measurements, one has to account for the radioactive decay of each of the isotopes. Since the beam-induced activity is many orders of magnitude larger than the natural cosmogenic induced activity, any pre-existing or post-beam cosmogenic activation can be ignored.  The number of isotope atoms at any given time $t$ since the beam exposure ($t$ = 0) is given by the Bateman equation without source terms
\begin{linenomath*}
\begin{align}
    N_{m,n}(t)=\sum_{i=1}^n N_{m,i}(0)\left[ \prod_{j=i}^{n-1} \lambda_j b_{j,j+1} \times \sum_{j=i}^n \left( \frac {e^{-\lambda_jt}} {\prod_{\overset{k=i}{\underset{k \ne j}{}}}^n(\lambda_k - \lambda_j)} \right) \right]
\end{align}
\end{linenomath*}
where $N_{m,i}(t)$ is the number of atoms of the $i$th isotope in the decay chain at time $t$ according to the cross section model $m$,  $\lambda_j$ is the decay constant of the $j$th isotope, and $b_{j, j+1}$ is the branching ratio between the $j$ and $j+1$ isotopes. For primary isotopes that are only directly produced by the beam, the above equation simplifies to a single exponential decay, which was calculated analytically based on the known half-lives. For primary isotopes that are part of a decay chain, we used Geant4 to simulate the decay time profiles, in order to automatically incorporate the correct branching ratios and half-lives for each of the isotopes. From the simulation it was found that, while several beam-produced isotopes contribute to the observed decay rate at the time of the measurement, the long time between beam exposure and measurement time meant that the decay time profile of all fit isotopes, except for $^{125m}$Te (discussed in Section~\ref{sec:analysis}), followed a single exponential decay with a half-life corresponding to a single long-lived isotope. For these decay chain estimates, the additional statistical uncertainty in determining the activity from the simulation was included in the overall uncertainty. The activity of the primary isotope due to the contribution of each individual feed isotope, calculated at the reference time of the first measurement, i.e.\ 18th Dec 2020 ($t$ = 411.12 days $\equiv T$), is shown in Table~\ref{tab:pred_act}, along with the sum of all the contributions.
As can be seen in the Table~\ref{tab:pred_act}, there are some isotopes (e.g. \isotope{Sb}{109}, \isotope{Xe}{121}) that have significant beam production but, according to all cross-section models, contribute less than 1\% to the primary fit isotope at the time of the measurement. Since our experimental data does not have sensitivity to these isotopes, they are not considered for further analysis. We note that our early HPGe measurement identified several primary isotopes (e.g.\isotope{In}{109}, \isotope{I}{121}) that can feed the observed long-lived isotopes. Those measurements (not reported here) will be used in the future to further constrain the activation models. 

\subsection{Comparison to Measurement}
To quantitatively compare our measured activities of isotopes in the fit to the predictions of each cross section model $m$, we calculate a scale factor $S_{m, i}$
\begin{align}
    S_{m, i} = \frac{F_i(T)}{\lambda_i N_{m,i}(T)}
\label{eq:scale}
\end{align}
where $F_i$ is the fit results for the activity of isotope $i$ at reference time $T$ (see Table~\ref{tab:CrystalActivity}). We note that this scale factor incorporates all the values and uncertainties associated with the experimental measurement, with the numerator including the counting and fit results and the denominator including the beam exposure and target crystals. The scale factor for each primary isotope is shown in Table~\ref{tab:scale_fact}. Since the time profile of the fit results for $^{125m}$Te tells us the activity of $^{125m}$Te produced directly by the beam separately from that produced by the decays of $^{125}$Sb and other parent isotopes (see Section~\ref{sec:analysis}), we have separated out the predictions for $^{125m}$Te and $^{125}$Sb.

It can be seen that the BERTINI model has the most accurate predictions (scale factor close to 1) for $^{22}$Na, and $^{125}$I, favoring the more recent experimental measurements of $^{22}$Na production by Uwamino et. al. \cite{uwamino1992measurement} and consistent with the comparison to the $^{125}$I experimental cross sections discussed in Section~\ref{sec:isotopes}. However, it severely underestimates the production rates for all metastable states (e.g. \isotope{Te}{123m}, \isotope{Te}{127m}). This behaviour appears to be a known issue for our version of Geant4 \cite{fermilab-bertini} and we therefore exclude the BERTINI predictions for all primary metastable isotopes in the remainder of this work.

\section{Cosmogenic Neutron Activation}
\label{sec:cosmogenic_rates}
\begin{table*}[t]
\begin{center}
\begin{tabular}{ccccccc}
\hline
\hline
Isotope  & \multicolumn{2}{c}{INCLXX}& \multicolumn{2}{c}{BERTINI}& \multicolumn{2}{c}{BIC}\\
 & Unscaled Prod. & Scaled Prod.  &  Unscaled Prod. & Scaled Prod & Unscaled Prod. & Scaled Prod. \\
& [atoms/kg/day] & [atoms/kg/day] & [atoms/kg/day] & [atoms/kg/day] & [atoms/kg/day] & [atoms/kg/day]\\
\hline
\vrule width 0pt height 2.2ex
$^3$H  &  \num{ 99 \pm 12 } & \num{ 76 \pm 15 } &  \num{ 32.1 \pm 4.0 } &  \num{ 96 \pm 19 }&  \num{ 57.4 \pm 7.2 } &  \num{ 69 \pm 12 } \\
$^{22}$Na  &  \num{ 117 \pm 15 } & \num{ 47.7 \pm 7.2 } &  \num{ 46.9 \pm 5.9 } &  \num{ 48.8 \pm 7.9 }&  \num{ 123 \pm 15 } &  \num{ 47.0 \pm 7.1 } \\
$^{109}$Cd  &  \num{ 3.54 \pm 0.44 } & \num{ 2.92 \pm 0.81 } &  \num{ 1.14 \pm 0.14 } &  \num{ 1.17 \pm 0.35 }&  \num{ 3.07 \pm 0.38 } &  \num{ 3.05 \pm 0.86 } \\
$^{109}$In  &  \num{ 2.34 \pm 0.29 } & \num{ 1.93 \pm 0.53 } &  \num{ 1.76 \pm 0.22 } &  \num{ 1.81 \pm 0.54 }&  \num{ 2.08 \pm 0.26 } &  \num{ 2.07 \pm 0.59 } \\
$^{109}$Sn  &  \num{ 0.57 \pm 0.07 } & \num{ 0.47 \pm 0.13 } &  \num{ 3.02 \pm 0.38 } &  \num{ 3.10 \pm 0.93 }&  \num{ 0.66 \pm 0.08 } &  \num{ 0.65 \pm 0.18 } \\
$^{113}$Sb  &  \num{ 1.24 \pm 0.16 } & \num{ 0.90 \pm 0.20 } &  \num{ 2.70 \pm 0.34 } &  \num{ 1.83 \pm 0.44 }&  \num{ 1.77 \pm 0.22 } &  \num{ 1.05 \pm 0.24 } \\
$^{113}$Sn  &  \num{ 1.40 \pm 0.17 } & \num{ 1.01 \pm 0.23 } &  \num{ 6.37 \pm 0.80 } &  \num{ 4.3 \pm 1.0 }&  \num{ 1.76 \pm 0.22 } &  \num{ 1.04 \pm 0.24 } \\
$^{113}$Te  &  $<0.01$ & $<0.01$ &  \num{ 1.56 \pm 0.20 } &  \num{ 1.06 \pm 0.25 } &  \num{ 0.015 \pm 0.002 } &  \num{ 0.009 \pm 0.002 } \\
$^{113m}$Sn  &  \num{ 7.72 \pm 0.96 } & \num{ 5.6 \pm 1.3 } & $<0.01$ & - &  \num{ 9.5 \pm 1.2 } &  \num{ 5.6 \pm 1.3 } \\
$^{121m}$Te  &  \num{ 25.0 \pm 3.1 } & \num{ 23.8 \pm 3.9 } & $<0.01$ & - &  \num{ 36.7 \pm 4.6 } &  \num{ 24.1 \pm 3.9 } \\
$^{123m}$Te  &  \num{ 28.7 \pm 3.6 } & \num{ 29.6 \pm 4.7 } & $<0.01$ & - &  \num{ 34.8 \pm 4.3 } &  \num{ 31.5 \pm 5.0 } \\
$^{125}$I  &  \num{ 160 \pm 20 } & \num{ 272 \pm 42 } &  \num{ 291 \pm 36 } &  \num{ 267 \pm 41 }&  \num{ 219 \pm 27 } &  \num{ 273 \pm 42 } \\
$^{125}$Sb  &  \num{ 7.76 \pm 0.97 } & \num{ 2.47 \pm 0.50 } &  \num{ 1.92 \pm 0.24 } &  \num{ 2.42 \pm 0.52 }&  \num{ 2.75 \pm 0.34 } &  \num{ 2.52 \pm 0.54 } \\
$^{125}$Sn  &  \num{ 0.29 \pm 0.04 } & \num{ 0.09 \pm 0.02 } &  \num{ 0.028 \pm 0.004 } &  \num{ 0.04 \pm 0.01 }&  \num{ 0.05 \pm 0.01 } &  \num{ 0.04 \pm 0.01 } \\
$^{125m}$Te  &  \num{ 17.5 \pm 2.2 } & \num{ 124 \pm 25 } &  $<0.01$  & - &  \num{ 13.6 \pm 1.7 } &  \num{ 127 \pm 25 } \\
$^{127m}$Te  &  \num{ 13.3 \pm 1.7 } & \num{ 5.4 \pm 4.2 } & $<0.01$ & - &  \num{ 4.45 \pm 0.56 } &  \num{ 5.8 \pm 4.5 } \\
\hline
\end{tabular}
\end{center}
\caption{Predictions of activation rates for NaI crystals exposed to the Gordon cosmogenic neutron flux \cite{gordon2004measurement} for each of the cross-section models considered. The first column for each model shows the rates directly calculated from the model while the second column shows the rates multiplied by the scale factor derived from comparing the model to the measurement made with the LANSCE beam.}
\label{tab:cosmo_rates}
\end{table*}

Having evaluated the scaling factor for each isotope, we may now compute their rates of cosmogenic neutron activation at sea level. The production rate of a given isotope in a sodium iodide crystal, $P_{m,i}$, in units of \mbox{\si{\atoms\per\kg\per\second}} is given by
\begin{linenomath*}
\begin{align}
\label{eq:production}
P_{m,i} = \sum_{x= Na, I}{n_x \int \Phi(E) \cdot \sigma_{m,i,x}(E)~dE},
\end{align}
\end{linenomath*}
where $m$ is again the assumed cross section model and $x$ is either \isotope{Na}{23} or \isotope{I}{127}. Here, $n$ is the number of target atoms per unit mass of NaI [atoms/kg], $\Phi(E)$ is the cosmic-ray neutron flux [\si{\neutrons\per\cm\squared\per\second\per\MeV}], and $\sigma$ is the production cross section [\si{\cm\squared}]. The integral is evaluated from 1\,MeV to 10\,GeV, with the lower bound set by the typical nuclear reaction threshold and the upper bound determined by the negligible flux of cosmic-ray neutrons above this energy.

There have been several measurements and calculations of the cosmic-ray neutron flux (see, e.g., Refs.~\cite{hess1959cosmic, armstrong1973calculations, ziegler1996terrestrial}). The intensity of the neutron flux varies with altitude, location in the geomagnetic field, and solar magnetic activity---though the spectral shape does not vary as significantly---and correction factors must be applied to calculate the appropriate flux \cite{desilets2001scaling}. The most commonly used reference spectrum for sea-level cosmic-ray neutrons is the so-called ``Gordon'' spectrum \cite{gordon2004measurement} (shown in Fig.~\ref{fig:fluence}), which is based on measurements at five different sites in the United States, scaled to sea level at the location of New York City during the mid-point of solar modulation. We used the parameterization given in Ref.~\cite{gordon2004measurement}, which agrees with the data to within a few percent. The spectrum uncertainties at high energies are dominated by uncertainties in the spectrometer detector response function ($<4$\% below 10 MeV and 10--15\% above 150 MeV). We have assigned an average uncertainty of 12.5\% across the entire energy range.

The predicted production rates for the cross-section models considered are shown in the unscaled columns of Table~\ref{tab:cosmo_rates}. While the cross section is not experimentally known across the entire energy range and each of the models predicts a different energy dependence, the similar shapes of the LANSCE beam and the cosmic-ray neutron spectrum allow us to greatly reduce the systematic uncertainty arising from the cross sections.  We obtain our best estimates for the neutron-induced cosmogenic production rate for a given isotope and cross section model by multiplying the production rate given in Eq.~\ref{eq:production} by its corresponding scale factor (obtained from the comparison of the model predictions to the measurements on the LANSCE beam)
\begin{align}
    P_{m,i}' = P_{m,i} \cdot S_{m,i},
\end{align}
The resultant values are shown in the second column under each cross-section model heading in Table~\ref{tab:cosmo_rates}. 

\begin{table*}[t]
\begin{center}
\begin{tabular}{ccccccc}
\hline
\hline
Isotope  & Cosmic Neutron & Experimental & Cross Section & Cosmic Neutron & Total\\
 & Activation Rate  &  Uncertainty  & Uncertainty  &  Uncertainty  &  Uncertainty  \\
& [atoms/kg/day] & [\%] & [\%] & [\%] & [\%] \\
\hline
\vrule width 0pt height 2.2ex
$^{3}$H  &  \num{ 80 \pm 21 } &  \num{ 15 } &  \num{ 17 } &  \num{ 13 } &  \num{ 26 } \\
$^{22}$Na  &  \num{ 47.8 \pm 7.8 } &  \num{ 10 } &  \num{ 2.0 } &  \num{ 13 } &  \num{ 16 } \\
$^{109}$Cd (+ $^{109}$In + $^{109}$Sn)   &  \num{ 5.7 \pm 1.8 } &  \num{ 27 } &  \num{ 6.6 } &  \num{ 13 } &  \num{ 31 } \\
$^{113}$Sn (+ $^{113m}$Sn + $^{113}$Sb + $^{113}$Te) &  \num{ 7.1 \pm 1.7 } &  \num{ 20 } &  \num{ 1.9 } &  \num{ 13 } &  \num{ 24 } \\
$^{121m}$Te  &  \num{ 24.0 \pm 3.9 } &  \num{ 10 } &  \num{ 0.95 } &  \num{ 13 } &  \num{ 16 } \\
$^{123m}$Te  &  \num{ 30.5 \pm 5.0 } &  \num{ 9.8 } &  \num{ 4.4 } &  \num{ 13 } &  \num{ 16 } \\
$^{125}$I  &  \num{ 271 \pm 42 } &  \num{ 9.2 } &  \num{ 1.2 } &  \num{ 13 } &  \num{ 16 } \\
$^{125}$Sb (+ $^{125}$Sn)  &  \num{ 2.53 \pm 0.55 } &  \num{ 18 } &  \num{ 2.6 } &  \num{ 13 } &  \num{ 22 } \\
$^{125m}$Te  &  \num{ 125 \pm 25 } &  \num{ 16 } &  \num{ 1.7 } &  \num{ 13 } &  \num{ 20 } \\
$^{127m}$Te  &  \num{ 5.6 \pm 4.4 } &  \num{ 76 } &  \num{ 4.6 } &  \num{ 13 } &  \num{ 77 } \\
\hline
\end{tabular}
\end{center}
\caption{Final cosmogenic neutron activation rates for isotopes in NaI crystals. Isotopes in parenthesis are short-lived isotopes whose decays feed the primary isotope and whose contributions, with the appropriate branching ratio, are included in the calculated activation rate. The third, fourth, and fifth columns show the contributions of the different components to the overall uncertainty shown in the sixth column.}
\label{tab:final_result}
\end{table*}

The spread in the values for the different cross-section models is an indication of the systematic uncertainty in the extrapolation from the LANSCE beam measurement to the cosmic-ray neutron spectrum. If the LANSCE neutron-beam spectral shape was the same as that of the cosmic-ray neutrons, or if the cross-section models all agreed in shape, the central values of the scaled production rates would be identical. We therefore use the average of the central values across all cross-section models as our combined central value and the standard deviation of the central values as our estimate of the uncertainty due to the cross-section shape, which is then combined with the other uncertainties that arise from our experimental measurements and the cosmic ray neutron flux. Table~\ref{tab:final_result} shows the final central values and uncertainties for the cosmic neutron activation rates, along with the individual contributions to the total uncertainties. For ease of comparison with previous measurements and use by future experiments, we have combined the contributions of short-lived (T$_{1/2} < 10$ days) isotopes (including the relevant branching ratios) that feed the primary long-lived isotope of relevance for dark matter searches.

\section{Discussion and Summary}
\label{sec:discussion}

\begin{figure}
\begin{center}
 \includegraphics[width=\columnwidth]{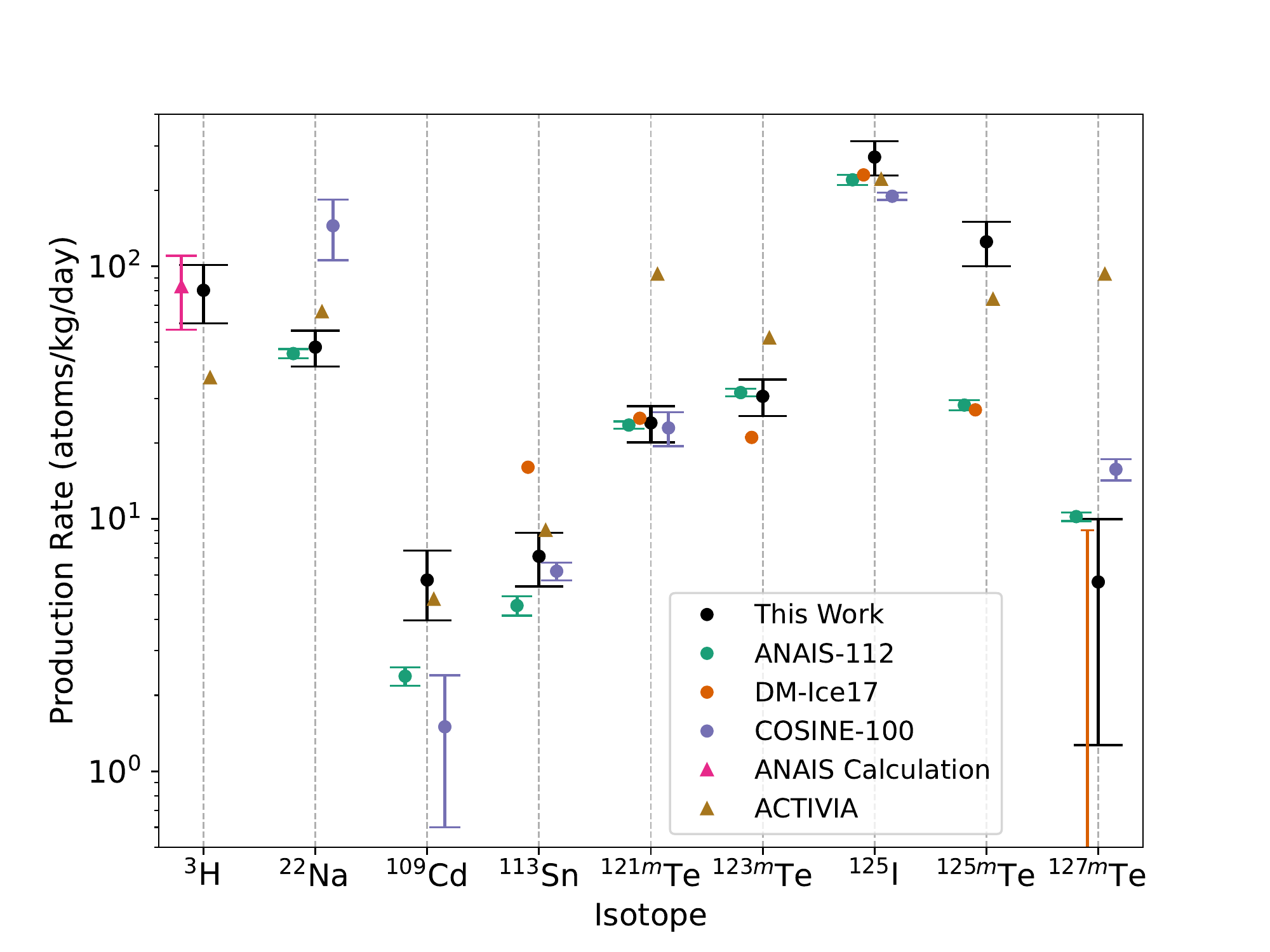}
 \end{center}
 \caption{Comparison of the results from this work (black circles) to previously measured comsogenic production rates (colored circles) in \mbox{ANAIS-112} \cite{anais2018activationrates, Amare:2018ndh}, \mbox{DM-Ice17} \cite{dmice17activationrates}, and \mbox{COSINE-100} \cite{COSINE100activationrates}. Also shown (colored triangles) are averaged calculations of tritium production from ANAIS \cite{anais2018activationrates} and estimates from the \mbox{ACTIVIA} \cite{back2008activia, anais2018activationrates, activia2016} semi-empirical model. See Sec.~\ref{sec:discussion} for details.}
 \label{fig:LitCompare}
\end{figure}

The cosmogenic production rates for several of the isotopes reported in Table \ref{tab:final_result} have been reported previously using estimates of the ambient cosmic ray exposure of NaI crystals above-ground, before measurement underground in a low-background environment \cite{anais2018activationrates, Amare:2018ndh,  dmice17activationrates, COSINE100activationrates}. A comparison between these measurements and our results are given in Figure~\ref{fig:LitCompare}. Values reported for the \mbox{ANAIS-112} experiment use the weighted average of their results across all crystals and the \mbox{COSINE-100} values shown use the weighted sum of all crystals reported except for crystals 3 and 4 which have complicated exposure histories. No uncertainties are reported for the measurements from the \mbox{DM-Ice17} experiment.

Overall, we obtain good agreement with the previously reported values for most isotopes. This is noteworthy because our methodology differs significantly from the previous approaches and is therefore subject to quite different experimental systematics. Our cosmogenic production rates only account for the production from neutron interactions, while ambient cosmic ray exposure will contain small contributions from other high energy particles, notably protons and muons. The contribution from protons is expected to be $\lesssim 10\%$ relative to that from neutrons \cite{anais2018activationrates} and the contribution from muons is expected to be even smaller.

As seen in Figure~\ref{fig:LitCompare}, our measurements agree, within uncertainties, with all the isotopes previously measured by \mbox{ANAIS-112} \cite{anais2018activationrates, Amare:2018ndh} except for \isotope{Te}{125m}. They report a \isotope{Te}{125m} production rate of \SI{28.2 \pm 1.3}{atoms/kg/day} compared to our measurement of \SI{125 \pm 25}{atoms/kg/day}. For both measurements, the estimate is driven by the time evolution of the peak at \SI{145}{\keV} (see Figures~\ref{fig:SimultaneousSpectra} and \ref{fig:Te125mDecay}), which is found to be consistent with the known \isotope{Te}{125m} half-life (in our case after subtracting the contribution of \isotope{Sb}{125}). The origin of the discrepancy is not understood. 

Our measured $^{22}$Na production rate is in good agreement with that reported by ANAIS-112 but both measurements differ from the value reported by COSINE-100. We note that isotopes of elements other than sodium and iodine present in the starting NaI powder are typically removed during the crystal growth process. However $^{22}$Na is chemically identical to the stable Na isotope and thus the amount present in the crystal will also depend upon the history of the NaI powder, not just the post-growth crystal exposure. We speculate that the discrepancy in the COSINE-100 results may be due to the exposure of the NaI powder prior to crystal manufacture that was not accounted for. One would expect similar differences in the estimated production rates of \isotope{I}{125}, though the shorter half-life mitigates the effect.

Figure~\ref{fig:LitCompare} also shows predictions from the semi-empirical cosmogenic activation code ACTIVIA \cite{back2008activia}, as reported in \cite{activia2016, anais2018activationrates}, which is often used to predict the cosmogenic activation of low-background experiments. The ACTIVIA results in general disagree with our measurements, and given the consistency amongst experimental results, we recommend that the experimental values be used to estimate the activation rate of isotopes that may be critical backgrounds.

Our results include the first experimental determination of the cosmogenic neutron activation rate for \trit. It agrees remarkably well with the analytical calculations of \trit~production from cosmogenic neutrons in Reference \cite{anais2018activationrates} where several different cross-section models were considered in different energy regimes and integrated with the same Gordon cosmogenic neutron spectrum used in this work (though they did not include the corresponding systematic uncertainty in the neutron spectrum). The overall range of all such calculations was used to estimate a value of \SI{83 \pm 27}{atoms/kg/day}. Our measured tritium production rate yields activities that are in agreement with the observed tritium activities seen in the COSINE and ANAIS experiments, given their best estimates for their NaI crystal exposure history. 
\section{Acknowledgements}
We would like to thank Mital Zalavadia for initial feasibility measurements and help with crystal procurement, Gabe Ortega for the design of the NaI crystal holders on the beam line, and Kyungwon Kim of the Center for Underground Physics at the Institute for Basic Science for the PMT voltage divider used in the measurement. We are grateful to Frank Wilkinson at Alpha Spectra for help developing the custom NaI crystal encapsulation design and providing detail on the materials and dimensions. We also thank Stephen Wender, Kranti Gunthoti, and Larry Rodriguez for technical assistance during the beam exposures. This research was supported in part by the Nuclear Physics, Particle Physics, Astro-Physics and Cosmology (NPAC) Initiative under the Laboratory Directed Research and Development Program at Pacific Northwest National Laboratory (PNNL), U.S.~National Science Foundation (NSF) Awards Nos.~PHY-1913742 and DGE-1122492, the Department of Energy, Office of Nuclear Physics under Federal Prime Agreements DE-FG02-01ER41166 and LANLEM78, by the Australian Government through the Australian Research Council Centre of Excellence for Dark Matter Particle Physics (CDM, CE200100008) and the Australian Research Council Discovery Program (DP170101675). Y.Y.~Zhong acknowledges stipend support from the Australian Research Council Centre of Excellence for Dark Matter Particle Physics. PNNL is a multiprogram national laboratory operated by Battelle for the U.S. Department of Energy.
\section{CR\MakeLowercase{edi}T authorship contribution statement}
\textbf{R.~Saldanha}: Conceptualization, Methodology, Validation, Formal analysis, Resources, Writing - Original Draft, Writing - Review \& Editing, Visualization, Supervision, Project administration, Funding acquisition.
\textbf{W.G.~Thompson}: Beam Exposure, Software, Validation, Formal analysis, Investigation, Data Curation, Writing - Original Draft, Writing - Review \& Editing.
\textbf{Y.Y.~Zhong}: Methodology, Software, Validation,  Formal analysis, Writing - Original Draft, Writing - Review \& Editing
\textbf{L.J.~Bignell}: Methodology, Software, Validation, Formal analysis, Writing - Original Draft, Writing - Review \& Editing, Visualization, Supervision, Funding acquisition.
\textbf{R.H.M.~Tsang}: Software, Validation, Formal Analysis, Writing - Original Draft, Writing - Review \& Editing, Visualization.
\textbf{S.J.~Hollick}: Investigation, Formal Analysis, Data Curation, Writing - Review \& Editing, Visualization.
\textbf{S.R.~Elliott}: Beam Exposure, Resources, Writing - Review \& Editing, Funding acquisition.
\textbf{G.J.~Lane}: Beam Exposure, Resources, Writing - Review \& Editing, Supervision, Funding acquisition.
\textbf{R.H.~Maruyama}: Resources, Supervision, Funding acquisition.
\textbf{L.~Yang}: Beam Exposure, Writing - Review \& Editing, Funding acquisition.

\bibliography{references_cap}

\begin{thebibliography}{61}
\expandafter\ifx\csname natexlab\endcsname\relax\def\natexlab#1{#1}\fi
\providecommand{\url}[1]{\texttt{#1}}
\providecommand{\href}[2]{#2}
\providecommand{\path}[1]{#1}
\providecommand{\DOIprefix}{doi:}
\providecommand{\ArXivprefix}{arXiv:}
\providecommand{\URLprefix}{URL: }
\providecommand{\Pubmedprefix}{pmid:}
\providecommand{\doi}[1]{\href{http://dx.doi.org/#1}{\path{#1}}}
\providecommand{\Pubmed}[1]{\href{pmid:#1}{\path{#1}}}
\providecommand{\bibinfo}[2]{#2}
\ifx\xfnm\relax \def\xfnm[#1]{\unskip,\space#1}\fi
\bibitem[{Knoll(2010)}]{knoll}
\bibinfo{author}{G.~F. Knoll}, \bibinfo{title}{Radiation detection and
  measurement}, \bibinfo{edition}{4th} ed., \bibinfo{publisher}{John Wiley},
  \bibinfo{address}{Hoboken, N.J}, \bibinfo{year}{2010}. \bibinfo{note}{OCLC:
  ocn612350364}.
\bibitem[{Bernabei et~al.(2018)Bernabei, , Belli, Bussolotti, Cappella,
  Caracciolo, Cerulli, Dai, d'Angelo, Marco, He, Incicchitti, Ma, Mattei,
  Merlo, Montecchia, Sheng, and Ye}]{DAMA_2018}
\bibinfo{author}{R.~Bernabei}, , \bibinfo{author}{P.~Belli},
  \bibinfo{author}{A.~Bussolotti}, \bibinfo{author}{F.~Cappella},
  \bibinfo{author}{V.~Caracciolo}, \bibinfo{author}{R.~Cerulli},
  \bibinfo{author}{C.~Dai}, \bibinfo{author}{A.~d'Angelo},
  \bibinfo{author}{A.~D. Marco}, \bibinfo{author}{H.~He},
  \bibinfo{author}{A.~Incicchitti}, \bibinfo{author}{X.~Ma},
  \bibinfo{author}{A.~Mattei}, \bibinfo{author}{V.~Merlo},
  \bibinfo{author}{F.~Montecchia}, \bibinfo{author}{X.~Sheng},
  \bibinfo{author}{Z.~Ye},
\newblock \bibinfo{title}{First model independent results from
  {DAMA}/{LIBRA}-phase2},
\newblock \bibinfo{journal}{Nuclear Physics and Atomic Energy}
  \bibinfo{volume}{19} (\bibinfo{year}{2018}) \bibinfo{pages}{307--325}.
\bibitem[{Drukier et~al.(1986)Drukier, Freese, and Spergel}]{Drukier_1986}
\bibinfo{author}{A.~K. Drukier}, \bibinfo{author}{K.~Freese},
  \bibinfo{author}{D.~N. Spergel},
\newblock \bibinfo{title}{Detecting cold dark-matter candidates},
\newblock \bibinfo{journal}{Phys. Rev. D} \bibinfo{volume}{33}
  (\bibinfo{year}{1986}) \bibinfo{pages}{3495--3508}.
\bibitem[{Bernabei et~al.(2021)}]{DAMA_2021}
\bibinfo{author}{R.~Bernabei}, et~al.,
\newblock \bibinfo{title}{{Further results from DAMA/Libra-phase2 and
  perspectives}},
\newblock \bibinfo{journal}{Nucl. Phys. Atom. Energy} \bibinfo{volume}{22}
  (\bibinfo{year}{2021}) \bibinfo{pages}{329--342}.
\bibitem[{Barbosa~de Souza et~al.(2017)}]{dmice_prd}
\bibinfo{author}{E.~Barbosa~de Souza}, et~al.
  (\bibinfo{collaboration}{DM-Ice}),
\newblock \bibinfo{title}{{First search for a dark matter annual modulation
  signal with NaI(Tl) in the Southern Hemisphere by DM-Ice17}},
\newblock \bibinfo{journal}{Phys. Rev. D} \bibinfo{volume}{95}
  (\bibinfo{year}{2017}) \bibinfo{pages}{032006}.
\bibitem[{Adhikari et~al.(2022)Adhikari, Barbosa~de Souza, Carlin, Choi, Choi,
  Ezeribe, Fran\ifmmode~\mbox{\c{c}}\else \c{c}\fi{}a, Ha, Hahn, Hollick, Jeon,
  Jo, Joo, Kang, Kauer, Kim, Kim, Kim, Kim, Kim, Kim, Kim, Kim, Kim, Ko, Kwon,
  Lee, Lee, Lee, Lee, Lee, Lee, Lee, Lee, Lee, Lee, Lee, Leonard, Manzato,
  Maruyama, Neal, Park, Park, Park, Park, Park, Pitta, Prihtiadi, Ra, Rott,
  Shin, Scarff, Spooner, Thompson, Yang, and Yu}]{cosine_3yr}
\bibinfo{author}{G.~Adhikari}, \bibinfo{author}{E.~Barbosa~de Souza},
  \bibinfo{author}{N.~Carlin}, \bibinfo{author}{J.~J. Choi},
  \bibinfo{author}{S.~Choi}, \bibinfo{author}{A.~C. Ezeribe},
  \bibinfo{author}{L.~E. Fran\ifmmode~\mbox{\c{c}}\else \c{c}\fi{}a},
  \bibinfo{author}{C.~Ha}, \bibinfo{author}{I.~S. Hahn}, \bibinfo{author}{S.~J.
  Hollick}, \bibinfo{author}{E.~J. Jeon}, \bibinfo{author}{J.~H. Jo},
  \bibinfo{author}{H.~W. Joo}, \bibinfo{author}{W.~G. Kang},
  \bibinfo{author}{M.~Kauer}, \bibinfo{author}{H.~Kim}, \bibinfo{author}{H.~J.
  Kim}, \bibinfo{author}{J.~Kim}, \bibinfo{author}{K.~W. Kim},
  \bibinfo{author}{S.~H. Kim}, \bibinfo{author}{S.~K. Kim},
  \bibinfo{author}{W.~K. Kim}, \bibinfo{author}{Y.~D. Kim},
  \bibinfo{author}{Y.~H. Kim}, \bibinfo{author}{Y.~J. Ko},
  \bibinfo{author}{H.~J. Kwon}, \bibinfo{author}{D.~H. Lee},
  \bibinfo{author}{E.~K. Lee}, \bibinfo{author}{H.~Lee}, \bibinfo{author}{H.~S.
  Lee}, \bibinfo{author}{H.~Y. Lee}, \bibinfo{author}{I.~S. Lee},
  \bibinfo{author}{J.~Lee}, \bibinfo{author}{J.~Y. Lee}, \bibinfo{author}{M.~H.
  Lee}, \bibinfo{author}{S.~H. Lee}, \bibinfo{author}{S.~M. Lee},
  \bibinfo{author}{D.~S. Leonard}, \bibinfo{author}{B.~B. Manzato},
  \bibinfo{author}{R.~H. Maruyama}, \bibinfo{author}{R.~J. Neal},
  \bibinfo{author}{B.~J. Park}, \bibinfo{author}{H.~K. Park},
  \bibinfo{author}{H.~S. Park}, \bibinfo{author}{K.~S. Park},
  \bibinfo{author}{S.~D. Park}, \bibinfo{author}{R.~L.~C. Pitta},
  \bibinfo{author}{H.~Prihtiadi}, \bibinfo{author}{S.~J. Ra},
  \bibinfo{author}{C.~Rott}, \bibinfo{author}{K.~A. Shin},
  \bibinfo{author}{A.~Scarff}, \bibinfo{author}{N.~J.~C. Spooner},
  \bibinfo{author}{W.~G. Thompson}, \bibinfo{author}{L.~Yang},
  \bibinfo{author}{G.~H. Yu} (\bibinfo{collaboration}{COSINE-100
  Collaboration}),
\newblock \bibinfo{title}{Three-year annual modulation search with cosine-100},
\newblock \bibinfo{journal}{Phys. Rev. D} \bibinfo{volume}{106}
  (\bibinfo{year}{2022}) \bibinfo{pages}{052005}.
\bibitem[{Thompson(2022)}]{thesis_thompson}
\bibinfo{author}{W.~G. Thompson}, \bibinfo{title}{Searching for Dark Matter
  with {COSINE-100}}, Ph.D. thesis, Yale University, \bibinfo{address}{New
  Haven, Connecticut, USA}, \bibinfo{year}{2022}.
\bibitem[{Amar\'e et~al.(2021)Amar\'e, Cebri\'an, Cintas, Coarasa, Garc\'{\i}a,
  Mart\'{\i}nez, Oliv\'an, Ortigoza, de~Sol\'orzano, Puimed\'on, Salinas,
  Sarsa, and Villar}]{anais3yrmodulation}
\bibinfo{author}{J.~Amar\'e}, \bibinfo{author}{S.~Cebri\'an},
  \bibinfo{author}{D.~Cintas}, \bibinfo{author}{I.~Coarasa},
  \bibinfo{author}{E.~Garc\'{\i}a}, \bibinfo{author}{M.~Mart\'{\i}nez},
  \bibinfo{author}{M.~A. Oliv\'an}, \bibinfo{author}{Y.~Ortigoza},
  \bibinfo{author}{A.~O. de~Sol\'orzano}, \bibinfo{author}{J.~Puimed\'on},
  \bibinfo{author}{A.~Salinas}, \bibinfo{author}{M.~L. Sarsa},
  \bibinfo{author}{P.~Villar},
\newblock \bibinfo{title}{{Annual modulation results from three-year exposure
  of ANAIS-112}},
\newblock \bibinfo{journal}{Phys. Rev. D} \bibinfo{volume}{103}
  (\bibinfo{year}{2021}) \bibinfo{pages}{102005}.
\bibitem[{Antonello et~al.(2019)Antonello, Barberio, Baroncelli, Benziger,
  Bignell, Bolognino, Calaprice, Copello, D'Angelo, D'Imperio, Dafinei, Carlo,
  Diemoz, Ludovico, Dix, Duffy, Froborg, Giovanetti, Hoppe, Ianni, Ioannucci,
  Krishnan, Lane, Mahmood, Mariani, Mastrodicasa, Montini, Mould, Nuti,
  Orlandi, Paris, Pettinacci, Pietrofaccia, Prokopovic, Rahatlou, Rossi,
  Sarbutt, Shields, Souza, Stuchbery, Suerfu, Tomei, Toso, Urquijo, Vignoli,
  Wada, Wallner, Williams, and Xu}]{Antonello_2019}
\bibinfo{author}{M.~Antonello}, \bibinfo{author}{E.~Barberio},
  \bibinfo{author}{T.~Baroncelli}, \bibinfo{author}{J.~Benziger},
  \bibinfo{author}{L.~J. Bignell}, \bibinfo{author}{I.~Bolognino},
  \bibinfo{author}{F.~Calaprice}, \bibinfo{author}{S.~Copello},
  \bibinfo{author}{D.~D'Angelo}, \bibinfo{author}{G.~D'Imperio},
  \bibinfo{author}{I.~Dafinei}, \bibinfo{author}{G.~D. Carlo},
  \bibinfo{author}{M.~Diemoz}, \bibinfo{author}{A.~D. Ludovico},
  \bibinfo{author}{W.~Dix}, \bibinfo{author}{A.~R. Duffy},
  \bibinfo{author}{F.~Froborg}, \bibinfo{author}{G.~K. Giovanetti},
  \bibinfo{author}{E.~Hoppe}, \bibinfo{author}{A.~Ianni},
  \bibinfo{author}{L.~Ioannucci}, \bibinfo{author}{S.~Krishnan},
  \bibinfo{author}{G.~J. Lane}, \bibinfo{author}{I.~Mahmood},
  \bibinfo{author}{A.~Mariani}, \bibinfo{author}{M.~Mastrodicasa},
  \bibinfo{author}{P.~Montini}, \bibinfo{author}{J.~Mould},
  \bibinfo{author}{F.~Nuti}, \bibinfo{author}{D.~Orlandi},
  \bibinfo{author}{M.~Paris}, \bibinfo{author}{V.~Pettinacci},
  \bibinfo{author}{L.~Pietrofaccia}, \bibinfo{author}{D.~Prokopovic},
  \bibinfo{author}{S.~Rahatlou}, \bibinfo{author}{N.~Rossi},
  \bibinfo{author}{A.~Sarbutt}, \bibinfo{author}{E.~Shields},
  \bibinfo{author}{M.~J. Souza}, \bibinfo{author}{A.~E. Stuchbery},
  \bibinfo{author}{B.~Suerfu}, \bibinfo{author}{C.~Tomei},
  \bibinfo{author}{V.~Toso}, \bibinfo{author}{P.~Urquijo},
  \bibinfo{author}{C.~Vignoli}, \bibinfo{author}{M.~Wada},
  \bibinfo{author}{A.~Wallner}, \bibinfo{author}{A.~G. Williams},
  \bibinfo{author}{J.~Xu},
\newblock \bibinfo{title}{The {SABRE} project and the {SABRE}
  proof-of-principle},
\newblock \bibinfo{journal}{The European Physical Journal C}
  \bibinfo{volume}{79} (\bibinfo{year}{2019}).
\bibitem[{Barberio et~al.(2022)Barberio, Baroncelli, Bignell, Bolognino,
  Brooks, Dastgiri, Duffy, Froehlich, Fu, Gerathy, Hill, Krishnan, Lane,
  Lawrence, Leaver, Mahmood, McGee, McKie, McNamara, Mews, Melbourne, Milana,
  Milligan, Mould, Nuti, Scutti, Slavkovská, Spinks, Stanley, Stuchbery,
  Taylor, Urquijo, Williams, Zhong, and Zurowski}]{SABRE_2022}
\bibinfo{author}{E.~Barberio}, \bibinfo{author}{T.~Baroncelli},
  \bibinfo{author}{L.~J. Bignell}, \bibinfo{author}{I.~Bolognino},
  \bibinfo{author}{G.~Brooks}, \bibinfo{author}{F.~Dastgiri},
  \bibinfo{author}{A.~R. Duffy}, \bibinfo{author}{M.~Froehlich},
  \bibinfo{author}{G.~Fu}, \bibinfo{author}{M.~S.~M. Gerathy},
  \bibinfo{author}{G.~C. Hill}, \bibinfo{author}{S.~Krishnan},
  \bibinfo{author}{G.~J. Lane}, \bibinfo{author}{G.~Lawrence},
  \bibinfo{author}{K.~T. Leaver}, \bibinfo{author}{I.~Mahmood},
  \bibinfo{author}{P.~McGee}, \bibinfo{author}{L.~McKie},
  \bibinfo{author}{P.~C. McNamara}, \bibinfo{author}{M.~Mews},
  \bibinfo{author}{W.~J.~D. Melbourne}, \bibinfo{author}{G.~Milana},
  \bibinfo{author}{L.~J. Milligan}, \bibinfo{author}{J.~Mould},
  \bibinfo{author}{F.~Nuti}, \bibinfo{author}{F.~Scutti},
  \bibinfo{author}{Z.~Slavkovská}, \bibinfo{author}{N.~J. Spinks},
  \bibinfo{author}{O.~Stanley}, \bibinfo{author}{A.~E. Stuchbery},
  \bibinfo{author}{G.~N. Taylor}, \bibinfo{author}{P.~Urquijo},
  \bibinfo{author}{A.~G. Williams}, \bibinfo{author}{Y.~Y. Zhong},
  \bibinfo{author}{M.~J. Zurowski}, \bibinfo{title}{Simulation of the {SABRE}
  {South} experiment and background characterization}, \bibinfo{year}{2022}.
  \URLprefix \url{https://arxiv.org/abs/2205.13849}.
  \DOIprefix\doi{10.48550/ARXIV.2205.13849}.
\bibitem[{Adhikari et~al.(2018)Adhikari, Adhikari, de~Souza, Carlin, Choi,
  Choi, Djamal, Ezeribe, Ha, Hahn, Hubbard, Jeon, Jo, Joo, Kang, Kang, Kauer,
  Kim, Kim, Kim, Kim, Kim, Kim, Kim, Kim, Kim, Kudryavtsev, Lee, Lee, Lee, Lee,
  Leonard, Lim, Lynch, Maruyama, Mouton, Olsen, Park, Park, Park, Park, Pettus,
  Pierpoint, Prihtiadi, Ra, Rogers, Rott, Scarff, Spooner, Thompson, Yang, and
  Yong}]{cosine_det}
\bibinfo{author}{G.~Adhikari}, \bibinfo{author}{P.~Adhikari},
  \bibinfo{author}{E.~B. de~Souza}, \bibinfo{author}{N.~Carlin},
  \bibinfo{author}{S.~Choi}, \bibinfo{author}{W.~Q. Choi},
  \bibinfo{author}{M.~Djamal}, \bibinfo{author}{A.~C. Ezeribe},
  \bibinfo{author}{C.~Ha}, \bibinfo{author}{I.~S. Hahn},
  \bibinfo{author}{A.~J.~F. Hubbard}, \bibinfo{author}{E.~J. Jeon},
  \bibinfo{author}{J.~H. Jo}, \bibinfo{author}{H.~W. Joo},
  \bibinfo{author}{W.~G. Kang}, \bibinfo{author}{W.~Kang},
  \bibinfo{author}{M.~Kauer}, \bibinfo{author}{B.~H. Kim},
  \bibinfo{author}{H.~Kim}, \bibinfo{author}{H.~J. Kim}, \bibinfo{author}{K.~W.
  Kim}, \bibinfo{author}{M.~C. Kim}, \bibinfo{author}{N.~Y. Kim},
  \bibinfo{author}{S.~K. Kim}, \bibinfo{author}{Y.~D. Kim},
  \bibinfo{author}{Y.~H. Kim}, \bibinfo{author}{V.~A. Kudryavtsev},
  \bibinfo{author}{H.~S. Lee}, \bibinfo{author}{J.~Lee}, \bibinfo{author}{J.~Y.
  Lee}, \bibinfo{author}{M.~H. Lee}, \bibinfo{author}{D.~S. Leonard},
  \bibinfo{author}{K.~E. Lim}, \bibinfo{author}{W.~A. Lynch},
  \bibinfo{author}{R.~H. Maruyama}, \bibinfo{author}{F.~Mouton},
  \bibinfo{author}{S.~L. Olsen}, \bibinfo{author}{H.~K. Park},
  \bibinfo{author}{H.~S. Park}, \bibinfo{author}{J.~S. Park},
  \bibinfo{author}{K.~S. Park}, \bibinfo{author}{W.~Pettus},
  \bibinfo{author}{Z.~P. Pierpoint}, \bibinfo{author}{H.~Prihtiadi},
  \bibinfo{author}{S.~Ra}, \bibinfo{author}{F.~R. Rogers},
  \bibinfo{author}{C.~Rott}, \bibinfo{author}{A.~Scarff},
  \bibinfo{author}{N.~J.~C. Spooner}, \bibinfo{author}{W.~G. Thompson},
  \bibinfo{author}{L.~Yang}, \bibinfo{author}{S.~H. Yong},
\newblock \bibinfo{title}{Initial performance of the {COSINE}-100 experiment},
\newblock \bibinfo{journal}{Eur. Phys. J. C} \bibinfo{volume}{78}
  (\bibinfo{year}{2018}) \bibinfo{pages}{107}.
\bibitem[{Antonello et~al.(2021)Antonello, Arnquist, Barberio, Baroncelli,
  Benziger, Bignell, Bolognino, Calaprice, Copello, Dafinei, D'Angelo,
  D'Imperio, D'Incecco, Carlo, Diemoz, Giacinto, Ludovico, Dix, Duffy, Hoppe,
  Ianni, Iannone, Ioannucci, Krishnan, Lane, Mahmood, Mariani, Milana, Mould,
  Nuti, Orlandi, Pettinacci, Pietrofaccia, Rahatlou, Scutti, Souza, Stuchbery,
  Suerfu, Tomei, Urquijo, Vignoli, Wallner, Wada, Williams, Zani, and
  Zurowski}]{Antonello_2021}
\bibinfo{author}{M.~Antonello}, \bibinfo{author}{I.~J. Arnquist},
  \bibinfo{author}{E.~Barberio}, \bibinfo{author}{T.~Baroncelli},
  \bibinfo{author}{J.~Benziger}, \bibinfo{author}{L.~J. Bignell},
  \bibinfo{author}{I.~Bolognino}, \bibinfo{author}{F.~Calaprice},
  \bibinfo{author}{S.~Copello}, \bibinfo{author}{I.~Dafinei},
  \bibinfo{author}{D.~D'Angelo}, \bibinfo{author}{G.~D'Imperio},
  \bibinfo{author}{M.~D'Incecco}, \bibinfo{author}{G.~D. Carlo},
  \bibinfo{author}{M.~Diemoz}, \bibinfo{author}{A.~D. Giacinto},
  \bibinfo{author}{A.~D. Ludovico}, \bibinfo{author}{W.~Dix},
  \bibinfo{author}{A.~R. Duffy}, \bibinfo{author}{E.~Hoppe},
  \bibinfo{author}{A.~Ianni}, \bibinfo{author}{M.~Iannone},
  \bibinfo{author}{L.~Ioannucci}, \bibinfo{author}{S.~Krishnan},
  \bibinfo{author}{G.~J. Lane}, \bibinfo{author}{I.~Mahmood},
  \bibinfo{author}{A.~Mariani}, \bibinfo{author}{S.~Milana},
  \bibinfo{author}{J.~Mould}, \bibinfo{author}{F.~Nuti},
  \bibinfo{author}{D.~Orlandi}, \bibinfo{author}{V.~Pettinacci},
  \bibinfo{author}{L.~Pietrofaccia}, \bibinfo{author}{S.~Rahatlou},
  \bibinfo{author}{F.~Scutti}, \bibinfo{author}{M.~Souza},
  \bibinfo{author}{A.~E. Stuchbery}, \bibinfo{author}{B.~Suerfu},
  \bibinfo{author}{C.~Tomei}, \bibinfo{author}{P.~Urquijo},
  \bibinfo{author}{C.~Vignoli}, \bibinfo{author}{A.~Wallner},
  \bibinfo{author}{M.~Wada}, \bibinfo{author}{A.~G. Williams},
  \bibinfo{author}{A.~Zani}, \bibinfo{author}{M.~Zurowski},
\newblock \bibinfo{title}{Characterization of {SABRE} crystal {NaI}-33 with
  direct underground counting},
\newblock \bibinfo{journal}{The European Physical Journal C}
  \bibinfo{volume}{81} (\bibinfo{year}{2021}).
\bibitem[{Villar et~al.(2018)Villar, Amar\'{e}
  et~al.}]{anais2018activationrates}
\bibinfo{author}{P.~Villar}, \bibinfo{author}{J.~Amar\'{e}}, et~al.,
\newblock \bibinfo{title}{{Study of the cosmogenic activation in NaI(Tl)
  crystals within the ANAIS experiment}},
\newblock \bibinfo{journal}{International Journal of Modern Physics A}
  \bibinfo{volume}{33} (\bibinfo{year}{2018}) \bibinfo{pages}{1843006}.
\bibitem[{{Barbosa de Souza} et~al.(2020)}]{COSINE100activationrates}
\bibinfo{author}{E.~{Barbosa de Souza}}, et~al.,
\newblock \bibinfo{title}{{Study of cosmogenic radionuclides in the COSINE-100
  NaI(Tl) detectors}},
\newblock \bibinfo{journal}{Astroparticle Physics} \bibinfo{volume}{115}
  (\bibinfo{year}{2020}) \bibinfo{pages}{102390}.
\bibitem[{Amar\'e et~al.(2015)}]{Amare:2014bea}
\bibinfo{author}{J.~Amar\'e}, et~al.,
\newblock \bibinfo{title}{{Cosmogenic radionuclide production in NaI(Tl)
  crystals}},
\newblock \bibinfo{journal}{JCAP} \bibinfo{volume}{02} (\bibinfo{year}{2015})
  \bibinfo{pages}{046}.
\bibitem[{Amare et~al.(2019)}]{Amare:2018ndh}
\bibinfo{author}{J.~Amare}, et~al.,
\newblock \bibinfo{title}{{Analysis of backgrounds for the ANAIS-112 dark
  matter experiment}},
\newblock \bibinfo{journal}{Eur. Phys. J. C} \bibinfo{volume}{79}
  (\bibinfo{year}{2019}) \bibinfo{pages}{412}.
\bibitem[{Pettus(2015)}]{dmice17activationrates}
\bibinfo{author}{W.~C. Pettus}, \bibinfo{title}{{Cosmogenic activation in NaI
  detectors for dark matter searches}}, Ph.D. thesis, Wisconsin U., Madison,
  Wisconsin U., Madison, \bibinfo{year}{2015}.
\bibitem[{Lisowski and Schoenberg(2006)}]{lisowski2006alamos}
\bibinfo{author}{P.~Lisowski}, \bibinfo{author}{K.~Schoenberg},
\newblock \bibinfo{title}{{The Los Alamos Neutron Science Center}},
\newblock \bibinfo{journal}{Nuclear Instruments and Methods A}
  \bibinfo{volume}{562} (\bibinfo{year}{2006}) \bibinfo{pages}{910--914}.
\bibitem[{Takala(2006)}]{icehouse}
\bibinfo{author}{B.~Takala},
\newblock \bibinfo{title}{{The ICE House}},
\newblock \bibinfo{journal}{Los Alamos Science}  (\bibinfo{year}{2006}).
\bibitem[{Saldanha et~al.(2019)Saldanha, Back, Tsang, Alexander, Elliott,
  Ferrara, Mace, Overman, and Zalavadia}]{Saldanha2019-Ar}
\bibinfo{author}{R.~Saldanha}, \bibinfo{author}{H.~O. Back},
  \bibinfo{author}{R.~H.~M. Tsang}, \bibinfo{author}{T.~Alexander},
  \bibinfo{author}{S.~R. Elliott}, \bibinfo{author}{S.~Ferrara},
  \bibinfo{author}{E.~Mace}, \bibinfo{author}{C.~Overman},
  \bibinfo{author}{M.~Zalavadia},
\newblock \bibinfo{title}{Cosmogenic production of $^{39}\mathrm{Ar}$ and
  $^{37}\mathrm{Ar}$ in argon},
\newblock \bibinfo{journal}{Phys. Rev. C} \bibinfo{volume}{100}
  (\bibinfo{year}{2019}) \bibinfo{pages}{024608}.
\bibitem[{Saldanha et~al.(2020)Saldanha, Thomas, Tsang, Chavarria, Bunker,
  Burnett, Elliott, Matalon, Mitra, Piers, Privitera, Ramanathan, and
  Smida}]{Saldanha2020-Si}
\bibinfo{author}{R.~Saldanha}, \bibinfo{author}{R.~Thomas},
  \bibinfo{author}{R.~H.~M. Tsang}, \bibinfo{author}{A.~E. Chavarria},
  \bibinfo{author}{R.~Bunker}, \bibinfo{author}{J.~L. Burnett},
  \bibinfo{author}{S.~R. Elliott}, \bibinfo{author}{A.~Matalon},
  \bibinfo{author}{P.~Mitra}, \bibinfo{author}{A.~Piers},
  \bibinfo{author}{P.~Privitera}, \bibinfo{author}{K.~Ramanathan},
  \bibinfo{author}{R.~Smida},
\newblock \bibinfo{title}{Cosmogenic activation of silicon},
\newblock \bibinfo{journal}{Phys. Rev. D} \bibinfo{volume}{102}
  (\bibinfo{year}{2020}) \bibinfo{pages}{102006}.
\bibitem[{Dunford and Burrows(1998)}]{dunford1998online}
\bibinfo{author}{C.~Dunford}, \bibinfo{author}{T.~Burrows},
\newblock \bibinfo{title}{Online nuclear data service,
  https://www.nndc.bnl.gov/}  (\bibinfo{year}{1998}).
\bibitem[{Qaim and W{\"o}lfle(1978)}]{qaim1978triton}
\bibinfo{author}{S.~Qaim}, \bibinfo{author}{R.~W{\"o}lfle},
\newblock \bibinfo{title}{{Triton emission in the interactions of fast neutrons
  with nuclei}},
\newblock \bibinfo{journal}{Nuclear Physics A} \bibinfo{volume}{295}
  (\bibinfo{year}{1978}) \bibinfo{pages}{150--162}.
\bibitem[{Boudard et~al.(2013)Boudard, Cugnon, David, Leray, and
  Mancusi}]{boudard2013new}
\bibinfo{author}{A.~Boudard}, \bibinfo{author}{J.~Cugnon},
  \bibinfo{author}{J.-C. David}, \bibinfo{author}{S.~Leray},
  \bibinfo{author}{D.~Mancusi},
\newblock \bibinfo{title}{{New potentialities of the Liege intranuclear cascade
  model for reactions induced by nucleons and light charged particles}},
\newblock \bibinfo{journal}{Physical Review C} \bibinfo{volume}{87}
  (\bibinfo{year}{2013}) \bibinfo{pages}{014606}.
\bibitem[{Mancusi et~al.(2014)Mancusi, Boudard, Cugnon, David, Kaitaniemi, and
  Leray}]{mancusi2014extension}
\bibinfo{author}{D.~Mancusi}, \bibinfo{author}{A.~Boudard},
  \bibinfo{author}{J.~Cugnon}, \bibinfo{author}{J.-C. David},
  \bibinfo{author}{P.~Kaitaniemi}, \bibinfo{author}{S.~Leray},
\newblock \bibinfo{title}{Extension of the {L}i{\`e}ge intranuclear-cascade
  model to reactions induced by light nuclei},
\newblock \bibinfo{journal}{Physical Review C} \bibinfo{volume}{90}
  (\bibinfo{year}{2014}) \bibinfo{pages}{054602}.
\bibitem[{Bertini(1963)}]{bertini1963low}
\bibinfo{author}{H.~W. Bertini},
\newblock \bibinfo{title}{{Low-energy intranuclear cascade calculation}},
\newblock \bibinfo{journal}{Physical Review} \bibinfo{volume}{131}
  (\bibinfo{year}{1963}) \bibinfo{pages}{1801}.
\bibitem[{Guthrie et~al.(1968)Guthrie, Alsmiller~Jr, and
  Bertini}]{guthrie1968calculation}
\bibinfo{author}{M.~Guthrie}, \bibinfo{author}{R.~Alsmiller~Jr},
  \bibinfo{author}{H.~Bertini},
\newblock \bibinfo{title}{{Calculation of the capture of negative pions in
  light elements and comparison with experiments pertaining to cancer
  radiotherapy}},
\newblock \bibinfo{journal}{Nuclear Instruments and Methods}
  \bibinfo{volume}{66} (\bibinfo{year}{1968}) \bibinfo{pages}{29--36}.
\bibitem[{Bertini(1969)}]{bertini1969intranuclear}
\bibinfo{author}{H.~W. Bertini},
\newblock \bibinfo{title}{{Intranuclear-cascade calculation of the secondary
  nucleon spectra from nucleon-nucleus interactions in the energy range 340 to
  2900 MeV and comparisons with experiment}},
\newblock \bibinfo{journal}{Physical Review} \bibinfo{volume}{188}
  (\bibinfo{year}{1969}) \bibinfo{pages}{1711}.
\bibitem[{Bertini and Guthrie(1971)}]{bertini1971news}
\bibinfo{author}{H.~W. Bertini}, \bibinfo{author}{M.~P. Guthrie},
\newblock \bibinfo{title}{{News item results from medium-energy
  intranuclear-cascade calculation}},
\newblock \bibinfo{journal}{Nuclear Physics A} \bibinfo{volume}{169}
  (\bibinfo{year}{1971}) \bibinfo{pages}{670--672}.
\bibitem[{Folger et~al.(2004)Folger, Ivanchenko, and
  Wellisch}]{folger2004binary}
\bibinfo{author}{G.~Folger}, \bibinfo{author}{V.~Ivanchenko},
  \bibinfo{author}{J.~Wellisch},
\newblock \bibinfo{title}{{The binary cascade}},
\newblock \bibinfo{journal}{The European Physical Journal A-Hadrons and Nuclei}
  \bibinfo{volume}{21} (\bibinfo{year}{2004}) \bibinfo{pages}{407--417}.
\bibitem[{Allison et~al.(2016)}]{allison2016recent}
\bibinfo{author}{J.~Allison}, et~al.,
\newblock \bibinfo{title}{{Recent developments in Geant4}},
\newblock \bibinfo{journal}{Nuclear Instruments and Methods A}
  \bibinfo{volume}{835} (\bibinfo{year}{2016}) \bibinfo{pages}{186--225}.
\bibitem[{Agostinelli et~al.(2003)}]{agostinelli2003geant4}
\bibinfo{author}{S.~Agostinelli}, et~al.,
\newblock \bibinfo{title}{{GEANT4---a simulation toolkit}},
\newblock \bibinfo{journal}{Nuclear Instruments and Methods A}
  \bibinfo{volume}{506} (\bibinfo{year}{2003}) \bibinfo{pages}{250--303}.
\bibitem[{Otuka et~al.(2014)Otuka, Dupont, Semkova, Pritychenko, Blokhin,
  Aikawa, Babykina, Bossant, Chen, Dunaeva, Forrest, Fukahori, Furutachi,
  Ganesan, Ge, Gritzay, Herman, Hlava{\v c}, Kat{\=o}, Lalremruata, Lee,
  Makinaga, Matsumoto, Mikhaylyukova, Pikulina, Pronyaev, Saxena, Schwerer,
  Simakov, Soppera, Suzuki, Tak{\'a}cs, Tao, Taova, T{\'a}rk{\'a}nyi, Varlamov,
  Wang, Yang, Zerkin, and Zhuang}]{exfor}
\bibinfo{author}{N.~Otuka}, \bibinfo{author}{E.~Dupont},
  \bibinfo{author}{V.~Semkova}, \bibinfo{author}{B.~Pritychenko},
  \bibinfo{author}{A.~Blokhin}, \bibinfo{author}{M.~Aikawa},
  \bibinfo{author}{S.~Babykina}, \bibinfo{author}{M.~Bossant},
  \bibinfo{author}{G.~Chen}, \bibinfo{author}{S.~Dunaeva},
  \bibinfo{author}{R.~Forrest}, \bibinfo{author}{T.~Fukahori},
  \bibinfo{author}{N.~Furutachi}, \bibinfo{author}{S.~Ganesan},
  \bibinfo{author}{Z.~Ge}, \bibinfo{author}{O.~Gritzay},
  \bibinfo{author}{M.~Herman}, \bibinfo{author}{S.~Hlava{\v c}},
  \bibinfo{author}{K.~Kat{\=o}}, \bibinfo{author}{B.~Lalremruata},
  \bibinfo{author}{Y.~Lee}, \bibinfo{author}{A.~Makinaga},
  \bibinfo{author}{K.~Matsumoto}, \bibinfo{author}{M.~Mikhaylyukova},
  \bibinfo{author}{G.~Pikulina}, \bibinfo{author}{V.~Pronyaev},
  \bibinfo{author}{A.~Saxena}, \bibinfo{author}{O.~Schwerer},
  \bibinfo{author}{S.~Simakov}, \bibinfo{author}{N.~Soppera},
  \bibinfo{author}{R.~Suzuki}, \bibinfo{author}{S.~Tak{\'a}cs},
  \bibinfo{author}{X.~Tao}, \bibinfo{author}{S.~Taova},
  \bibinfo{author}{F.~T{\'a}rk{\'a}nyi}, \bibinfo{author}{V.~Varlamov},
  \bibinfo{author}{J.~Wang}, \bibinfo{author}{S.~Yang},
  \bibinfo{author}{V.~Zerkin}, \bibinfo{author}{Y.~Zhuang},
\newblock \bibinfo{title}{{Towards a More Complete and Accurate Experimental
  Nuclear Reaction Data Library (EXFOR): International Collaboration Between
  Nuclear Reaction Data Centres (NRDC)}},
\newblock \bibinfo{journal}{Nuclear Data Sheets} \bibinfo{volume}{120}
  (\bibinfo{year}{2014}) \bibinfo{pages}{272 -- 276}.
\bibitem[{Liskien and Paulsen(1965)}]{liskien1965cross}
\bibinfo{author}{H.~Liskien}, \bibinfo{author}{A.~Paulsen},
\newblock \bibinfo{title}{{Cross-sections for the reactions Cu63 (n, $\alpha$)
  Co60, Ni60 (n, p) Co60, Ti46 (n, p) Sc46 and Na23 (n, 2n) Na22}},
\newblock \bibinfo{journal}{Nuclear Physics} \bibinfo{volume}{63}
  (\bibinfo{year}{1965}) \bibinfo{pages}{393--400}.
\bibitem[{Uwamino et~al.(1992)Uwamino, Sugita, Kondo, and
  Nakamura}]{uwamino1992measurement}
\bibinfo{author}{Y.~Uwamino}, \bibinfo{author}{H.~Sugita},
  \bibinfo{author}{Y.~Kondo}, \bibinfo{author}{T.~Nakamura},
\newblock \bibinfo{title}{{Measurement of neutron activation cross sections of
  energy up to 40 MeV using semimonoenergetic p-Be neutrons}},
\newblock \bibinfo{journal}{Nuclear science and engineering}
  \bibinfo{volume}{111} (\bibinfo{year}{1992}) \bibinfo{pages}{391--403}.
\bibitem[{Qaim and Ejaz(1968)}]{qaim1968half}
\bibinfo{author}{S.~Qaim}, \bibinfo{author}{M.~Ejaz},
\newblock \bibinfo{title}{{Half-lives and activation cross-sections of some
  radio-isotopes of iodine, tellurium and antimony formed in the interactions
  of iodine with 14.7 MeV neutrons}},
\newblock \bibinfo{journal}{Journal of Inorganic and Nuclear Chemistry}
  \bibinfo{volume}{30} (\bibinfo{year}{1968}) \bibinfo{pages}{2577--2581}.
\bibitem[{Liskien(1968)}]{liskien1968n}
\bibinfo{author}{H.~Liskien},
\newblock \bibinfo{title}{{(n, 3n) Processes and the Statistical Theory}},
\newblock \bibinfo{journal}{Nuclear Physics A} \bibinfo{volume}{118}
  (\bibinfo{year}{1968}) \bibinfo{pages}{379--388}.
\bibitem[{alp(2019)}]{alphaspectra}
\bibinfo{title}{{Alpha Spectra, Inc.}}, \bibinfo{year}{2019}. \URLprefix
  \url{https://alphaspectra.com/}.
\bibitem[{Kubota et~al.(1999)Kubota, Shiraishi, and Takami}]{Kubota1999NaIdmg}
\bibinfo{author}{S.~Kubota}, \bibinfo{author}{F.~Shiraishi},
  \bibinfo{author}{Y.~Takami},
\newblock \bibinfo{title}{{Radiation Damage of NaI(Tl) by Fast Neutron
  Irradiation: Blocking of the Energy Transfer Processes from V k Centers and
  Electrons to the Activator of Tl}},
\newblock \bibinfo{journal}{Journal of the Physical Society of Japan}
  \bibinfo{volume}{68} (\bibinfo{year}{1999}) \bibinfo{pages}{298--302}.
\bibitem[{Sudac and Valkovic(2010)}]{Sudac2010NaIDmg}
\bibinfo{author}{D.~Sudac}, \bibinfo{author}{V.~Valkovic},
\newblock \bibinfo{title}{{Irradiation of 4"×4" NaI(Tl) detector by the 14 MeV
  neutrons}},
\newblock \bibinfo{journal}{Applied Radiation and Isotopes}
  \bibinfo{volume}{68} (\bibinfo{year}{2010}) \bibinfo{pages}{896--900}.
  \bibinfo{note}{The 7th International Topical Meeting on Industrial Radiation
  and Radio isotope Measurement Application(IRRMA-7)}.
\bibitem[{Wender et~al.(1993)}]{wender1993fission}
\bibinfo{author}{S.~Wender}, et~al.,
\newblock \bibinfo{title}{{A fission ionization detector for neutron flux
  measurements at a spallation source}},
\newblock \bibinfo{journal}{Nuclear Instruments and Methods A}
  \bibinfo{volume}{336} (\bibinfo{year}{1993}) \bibinfo{pages}{226--231}.
\bibitem[{Gordon et~al.(2004)Gordon, Goldhagen, Rodbell, Zabel, Tang, Clem, and
  Bailey}]{gordon2004measurement}
\bibinfo{author}{M.~Gordon}, \bibinfo{author}{P.~Goldhagen},
  \bibinfo{author}{K.~Rodbell}, \bibinfo{author}{T.~Zabel},
  \bibinfo{author}{H.~Tang}, \bibinfo{author}{J.~Clem},
  \bibinfo{author}{P.~Bailey},
\newblock \bibinfo{title}{{Measurement of the flux and energy spectrum of
  cosmic-ray induced neutrons on the ground}},
\newblock \bibinfo{journal}{IEEE Transactions on Nuclear Science}
  \bibinfo{volume}{51} (\bibinfo{year}{2004}) \bibinfo{pages}{3427--3434}.
\bibitem[{Lisowski et~al.(1991)Lisowski, Gavron, Parker, Ullmann, Balestrini,
  Carlson, Wasson, and Hill}]{lisowski1991fission}
\bibinfo{author}{P.~Lisowski}, \bibinfo{author}{A.~Gavron},
  \bibinfo{author}{W.~Parker}, \bibinfo{author}{J.~Ullmann},
  \bibinfo{author}{S.~Balestrini}, \bibinfo{author}{A.~Carlson},
  \bibinfo{author}{O.~Wasson}, \bibinfo{author}{N.~Hill},
  \bibinfo{title}{{Fission cross sections in the intermediate energy region}},
  \bibinfo{type}{Technical Report}, Los Alamos National Lab., NM (USA),
  \bibinfo{year}{1991}.
\bibitem[{Carlson et~al.(2009)Carlson, Pronyaev, Smith, Larson, Chen, Hale,
  Hambsch, Gai, Oh, Badikov et~al.}]{carlson2009international}
\bibinfo{author}{A.~Carlson}, \bibinfo{author}{V.~Pronyaev},
  \bibinfo{author}{D.~Smith}, \bibinfo{author}{N.~M. Larson},
  \bibinfo{author}{Z.~Chen}, \bibinfo{author}{G.~Hale}, \bibinfo{author}{F.-J.
  Hambsch}, \bibinfo{author}{E.~Gai}, \bibinfo{author}{S.-Y. Oh},
  \bibinfo{author}{S.~Badikov}, et~al.,
\newblock \bibinfo{title}{{International evaluation of neutron cross section
  standards}},
\newblock \bibinfo{journal}{Nuclear Data Sheets} \bibinfo{volume}{110}
  (\bibinfo{year}{2009}) \bibinfo{pages}{3215--3324}.
\bibitem[{Tovesson et~al.(2014)Tovesson, Laptev, and Hill}]{tovesson2014fast}
\bibinfo{author}{F.~Tovesson}, \bibinfo{author}{A.~Laptev},
  \bibinfo{author}{T.~Hill},
\newblock \bibinfo{title}{{Fast neutron--induced fission cross sections of 233,
  234, 236, 238U up to 200 MeV}},
\newblock \bibinfo{journal}{Nuclear Science and Engineering}
  \bibinfo{volume}{178} (\bibinfo{year}{2014}) \bibinfo{pages}{57--65}.
\bibitem[{Marcinkevicius et~al.(2015)Marcinkevicius, Simakov, and
  Pronyaev}]{marcinkevicius2015209}
\bibinfo{author}{B.~Marcinkevicius}, \bibinfo{author}{S.~Simakov},
  \bibinfo{author}{V.~Pronyaev}, \bibinfo{title}{{209Bi (n, f) and natPb (n, f)
  Cross Sections as a New Reference and Extension of the 235U, 238U and 239Pu
  (n, f) Standards up to 1 GeV}}, \bibinfo{type}{Technical Report},
  International Atomic Energy Agency, \bibinfo{year}{2015}.
\bibitem[{Miller(2015)}]{miller2015measurement}
\bibinfo{author}{Z.~W. Miller}, \bibinfo{title}{{A measurement of the prompt
  fission neutron energy spectrum for 235U (n, f) and the neutron-induced
  fission cross section for 238U (n, f)}}, Ph.D. thesis, University of
  Kentucky, \bibinfo{year}{2015}.
\bibitem[{Duran et~al.(2017)Duran, Ventura, Meo, Tarr{\'\i}o, Tassan-Got, and
  Paradela}]{duran2017search}
\bibinfo{author}{I.~Duran}, \bibinfo{author}{A.~Ventura},
  \bibinfo{author}{S.~L. Meo}, \bibinfo{author}{D.~Tarr{\'\i}o},
  \bibinfo{author}{L.~Tassan-Got}, \bibinfo{author}{C.~Paradela},
\newblock \bibinfo{title}{{On the search for a (n, f) cross-section reference
  at intermediate energies}},
\newblock in: \bibinfo{booktitle}{{EPJ Web of Conferences}}, volume
  \bibinfo{volume}{146}, \bibinfo{organization}{EDP Sciences},
  \bibinfo{year}{2017}, p. \bibinfo{pages}{02032}.
\bibitem[{Carlson et~al.(2018)Carlson, Pronyaev, Capote, Hale, Chen, Duran,
  Hambsch, Kunieda, Mannhart, Marcinkevicius et~al.}]{carlson2018evaluation}
\bibinfo{author}{A.~Carlson}, \bibinfo{author}{V.~G. Pronyaev},
  \bibinfo{author}{R.~Capote}, \bibinfo{author}{G.~Hale},
  \bibinfo{author}{Z.-P. Chen}, \bibinfo{author}{I.~Duran},
  \bibinfo{author}{F.-J. Hambsch}, \bibinfo{author}{S.~Kunieda},
  \bibinfo{author}{W.~Mannhart}, \bibinfo{author}{B.~Marcinkevicius}, et~al.,
\newblock \bibinfo{title}{Evaluation of the neutron data standards},
\newblock \bibinfo{journal}{Nuclear Data Sheets} \bibinfo{volume}{148}
  (\bibinfo{year}{2018}) \bibinfo{pages}{143--188}.
\bibitem[{Rooney and Valentine(1997)}]{rooney_1997}
\bibinfo{author}{B.~Rooney}, \bibinfo{author}{J.~Valentine},
\newblock \bibinfo{title}{Scintillator light yield nonproportionality:
  calculating photon response using measured electron response},
\newblock \bibinfo{journal}{IEEE Transactions on Nuclear Science}
  \bibinfo{volume}{44} (\bibinfo{year}{1997}) \bibinfo{pages}{509--516}.
\bibitem[{Mengesha et~al.(1998)Mengesha, Taulbee, Rooney, and
  Valentine}]{mengesha_1998}
\bibinfo{author}{W.~Mengesha}, \bibinfo{author}{T.~Taulbee},
  \bibinfo{author}{B.~Rooney}, \bibinfo{author}{J.~Valentine},
\newblock \bibinfo{title}{Light yield nonproportionality of {CsI}({Tl}),
  {CsI}({Na}), and {YAP}},
\newblock \bibinfo{journal}{IEEE Transactions on Nuclear Science}
  \bibinfo{volume}{45} (\bibinfo{year}{1998}) \bibinfo{pages}{456--461}.
\bibitem[{Moses et~al.(2008)Moses, Payne, Choong, Hull, and
  Reutter}]{Moses2008}
\bibinfo{author}{W.~W. Moses}, \bibinfo{author}{S.~A. Payne},
  \bibinfo{author}{W.-S. Choong}, \bibinfo{author}{G.~Hull},
  \bibinfo{author}{B.~W. Reutter},
\newblock \bibinfo{title}{Scintillator non-proportionality: Present
  understanding and future challenges},
\newblock \bibinfo{journal}{IEEE Transactions on Nuclear Science}
  \bibinfo{volume}{55} (\bibinfo{year}{2008}) \bibinfo{pages}{1049--1053}.
\bibitem[{Payne et~al.(2011)Payne, Moses, Sheets, Ahle, Cherepy, Sturm,
  Dazeley, Bizarri, and Choong}]{payne2011nonproportionality}
\bibinfo{author}{S.~A. Payne}, \bibinfo{author}{W.~W. Moses},
  \bibinfo{author}{S.~Sheets}, \bibinfo{author}{L.~Ahle},
  \bibinfo{author}{N.~J. Cherepy}, \bibinfo{author}{B.~Sturm},
  \bibinfo{author}{S.~Dazeley}, \bibinfo{author}{G.~Bizarri},
  \bibinfo{author}{W.-S. Choong},
\newblock \bibinfo{title}{Nonproportionality of scintillator detectors: Theory
  and experiment. ii},
\newblock \bibinfo{journal}{IEEE Transactions on Nuclear Science}
  \bibinfo{volume}{58} (\bibinfo{year}{2011}) \bibinfo{pages}{3392--3402}.
\bibitem[{Robert and Casella(2005)}]{Robert2005}
\bibinfo{author}{C.~Robert}, \bibinfo{author}{G.~Casella},
  \bibinfo{title}{Monte Carlo Statistical Methods}, Springer Texts in
  Statistics, \bibinfo{edition}{2} ed., \bibinfo{publisher}{Springer},
  \bibinfo{address}{New York, NY}, \bibinfo{year}{2005}.
\bibitem[{Genser(2020)}]{fermilab-bertini}
\bibinfo{author}{K.~Genser}, \bibinfo{title}{{Recent evolution in Geant4}},
  \bibinfo{year}{2020}. \URLprefix
  \url{https://indico.fnal.gov/event/23998/#2-recent-evolution-in-geant4}.
\bibitem[{Hess et~al.(1959)Hess, Patterson, Wallace, and
  Chupp}]{hess1959cosmic}
\bibinfo{author}{W.~Hess}, \bibinfo{author}{H.~Patterson},
  \bibinfo{author}{R.~Wallace}, \bibinfo{author}{E.~Chupp},
\newblock \bibinfo{title}{Cosmic-ray neutron energy spectrum},
\newblock \bibinfo{journal}{Physical Review} \bibinfo{volume}{116}
  (\bibinfo{year}{1959}) \bibinfo{pages}{445}.
\bibitem[{Armstrong et~al.(1973)Armstrong, Chandler, and
  Barish}]{armstrong1973calculations}
\bibinfo{author}{T.~Armstrong}, \bibinfo{author}{K.~Chandler},
  \bibinfo{author}{J.~Barish},
\newblock \bibinfo{title}{Calculations of neutron flux spectra induced in the
  earth's atmosphere by galactic cosmic rays},
\newblock \bibinfo{journal}{Journal of Geophysical Research}
  \bibinfo{volume}{78} (\bibinfo{year}{1973}) \bibinfo{pages}{2715--2726}.
\bibitem[{Ziegler(1996)}]{ziegler1996terrestrial}
\bibinfo{author}{J.~Ziegler},
\newblock \bibinfo{title}{Terrestrial cosmic rays},
\newblock \bibinfo{journal}{IBM Journal of Research and Development}
  \bibinfo{volume}{40} (\bibinfo{year}{1996}) \bibinfo{pages}{19--39}.
\bibitem[{Desilets and Zreda(2001)}]{desilets2001scaling}
\bibinfo{author}{D.~Desilets}, \bibinfo{author}{M.~Zreda},
\newblock \bibinfo{title}{On scaling cosmogenic nuclide production rates for
  altitude and latitude using cosmic-ray measurements},
\newblock \bibinfo{journal}{Earth and Planetary Science Letters}
  \bibinfo{volume}{193} (\bibinfo{year}{2001}) \bibinfo{pages}{213--225}.
\bibitem[{Back and Ramachers(2008)}]{back2008activia}
\bibinfo{author}{J.~Back}, \bibinfo{author}{Y.~A. Ramachers},
\newblock \bibinfo{title}{{ACTIVIA: Calculation of isotope production
  cross-sections and yields}},
\newblock \bibinfo{journal}{Nuclear Instruments and Methods in Physics Research
  Section A: Accelerators, Spectrometers, Detectors and Associated Equipment}
  \bibinfo{volume}{586} (\bibinfo{year}{2008}) \bibinfo{pages}{286--294}.
\bibitem[{Zhang et~al.(2016)Zhang, Mei, Kudryavtsev, and
  Fiorucci}]{activia2016}
\bibinfo{author}{C.~Zhang}, \bibinfo{author}{D.-M. Mei},
  \bibinfo{author}{V.~Kudryavtsev}, \bibinfo{author}{S.~Fiorucci},
\newblock \bibinfo{title}{Cosmogenic activation of materials used in rare event
  search experiments},
\newblock \bibinfo{journal}{Astroparticle Physics} \bibinfo{volume}{84}
  (\bibinfo{year}{2016}) \bibinfo{pages}{62--69}.

\end{thebibliography}
\end{document}